\documentclass[10pt,prd,twocolumn,nofootinbib,notitlepage,aps,tightenlines,floatfix,
preprintnumbers,amsmath,amssymb,amsfonts,showpacs,superscriptaddress]{revtex4-2}

\RequirePackage[colorlinks=true
,urlcolor=blue
,anchorcolor=blue
,citecolor=blue
,filecolor=blue
,linkcolor=blue
,menucolor=blue
,linktocpage=true
,pdfproducer=medialab
,pdfa=true
]{hyperref}

\usepackage{graphicx,epstopdf}
\graphicspath{ {./Figures/} }
\usepackage{mathtools}
\usepackage{comment}
\usepackage{natbib,bm}
\usepackage[usenames,dvipsnames]{color}
\usepackage{slashed}    
\usepackage{xcolor}
\usepackage{multirow}
\usepackage{grffile}
\usepackage{physics}
\usepackage{dsfont}
\usepackage[]{hyperref}
\hypersetup{
	colorlinks,
	linkcolor={blue},
	linktocpage=true,
	citecolor={blue},
	urlcolor={blue}
}
\usepackage[capitalize]{cleveref}

\usepackage{mathtools, nccmath}
\usepackage{lipsum}
\usepackage{cancel}
\usepackage[normalem]{ulem}

\def\be{\begin{equation}}
\def\ee{\end{equation}}
\def\ba{\begin{eqnarray}}
\def\ea{\end{eqnarray}}
\def\ge{\mathrel{\raise.3ex\hbox{$>$\kern-.75em\lower1ex\hbox{$\sim$}}}}
\def\la{\mathrel{\raise.3ex\hbox{$<$\kern-.75em\lower1ex\hbox{$\sim$}}}}

\def\simgt{\mathrel{\raise.3ex\hbox{$>$\kern-.75em\lower1ex\hbox{$\sim$}}}}
\def\simlt{\mathrel{\raise.3ex\hbox{$<$\kern-.75em\lower1ex\hbox{$\sim$}}}}

\newcommand{\nc}{\newcommand}

\nc{\gone}{\bar g_{\pi NN}^{(1)}}
\nc{\gzero}{\bar g_{\pi NN}^{(0)}}
\nc{\al}{\alpha}
\nc{\ga}{\gamma}
\nc{\de}{\delta}
\nc{\ep}{\epsilon}
\nc{\ze}{\zeta}
\nc{\et}{\eta}
\nc{\ka}{\kappa}
\nc{\rh}{\rho}
\nc{\si}{\sigma}
\nc{\ta}{\tau}
\nc{\up}{\upsilon}
\nc{\ph}{\phi}
\nc{\ch}{\chi}
\nc{\ps}{\psi}
\nc{\om}{\omega}
\nc{\Ga}{\Gamma}
\nc{\De}{\Delta}
\nc{\La}{\Lambda}
\nc{\Si}{\Sigma}
\nc{\Up}{\Upsilon}
\nc{\Ph}{\Phi}
\nc{\Ps}{\Psi}
\nc{\Om}{\Omega}
\nc{\ptl}{\partial}
\nc{\del}{\nabla}
\nc{\ov}{\overline}
\nc{\newcaption}[1]{\centerline{\parbox{15cm}{\caption{#1}}}}
\nc{\us}{U(1)$_S$}

\def\beq{\begin{equation}}
\def\eeq{\end{equation}}
\def\bmat{\begin{displaymath}}
\def\emat{\end{displaymath}}
\def\bear{\begin{eqnarray}}
\def\eear{\end{eqnarray}}
\def\ba{\begin{eqnarray}}
\def\ea{\end{eqnarray}}
\def\bery{\begin{array}}
\def\ery{\end{array}}
\def\bit{\begin{itemize}}
\def\eit{\end{itemize}}
\def\ben{\begin{enumerate}}
\def\een{\end{enumerate}}
\def\btab{\begin{tabular}}
\def\etab{\end{tabular}}
\def\btbl{\begin{table}}
\def\etbl{\end{table}}
\def\bfig{\begin{figure}[htb]}
\def\efig{\end{figure}}
\def\bpic{\begin{picture}}
\def\epic{\end{picture}}

\def\ga{\mathrel{\raise.3ex\hbox{$>$\kern-.75em\lower1ex\hbox{$\sim$}}}}
\def\la{\mathrel{\raise.3ex\hbox{$<$\kern-.75em\lower1ex\hbox{$\sim$}}}}
\def\gappeq{\mathrel{\rlap {\raise.5ex\hbox{$>$}}
{\lower.5ex\hbox{$\sim$}}}}
\def\lappeq{\mathrel{\rlap{\raise.5ex\hbox{$<$}}
{\lower.5ex\hbox{$\sim$}}}}

\def\gyr{{\rm \, G\kern-0.125em yr}}
\def\mev{{\rm \, Me\kern-0.125em V}}
\def\gev{{\rm \, Ge\kern-0.125em V}}
\def\tev{{\rm \, Te\kern-0.125em V}}

\begin{document}

\title{BBN Constraints on the Hadronic Annihilation of sub-GeV Dark Matter}

\author{Afif Omar}
 \email{aomar@triumf.ca}
\affiliation{Department of Physics and Astronomy, University of Victoria, Victoria, BC V8P 5C2, Canada}
\affiliation{TRIUMF, 4004 Wesbrook Mall, Vancouver, BC V6T 2A3, Canada}
\author{Adam Ritz}
 \email{aritz@uvic.ca}
\affiliation{Department of Physics and Astronomy, University of Victoria, Victoria, BC V8P 5C2, Canada}

\date{October 14, 2025}

\begin{abstract}
\noindent 

We investigate the impact of residual annihilation from sub-GeV mass thermal relic dark matter candidates during big bang nucleosynthesis (BBN). Focusing on candidates with $p$-wave annihilation channels, we show that the hadronic injection of pions and kaons beyond freeze-out, and their subsequent interaction with protons and neutrons prior to the deuterium bottleneck, provides a sensitivity to annihilation that surpasses that of the CMB and indirect detection in the galaxy.

\end{abstract}
\maketitle

\vspace{-2mm}
\section{Introduction}

Models of thermal relic dark matter (DM) are characterized by an annihilation rate $\langle \si v\rangle$ that in the early Universe sets the time of freeze-out and thus the relic abundance. The fact that the observed dark matter density requires $\langle \si v\rangle \sim 1$~pb initially focused attention on the electroweak scale, but extensive efforts have been made in recent years to explore the full mass range for thermal relic DM including the sub-GeV regime. Models in this mass range are necessarily part of extended dark sectors with a new force mediator for annihilation due to the Lee-Weinberg bound~\cite{Lee:1977ua}. Such models are also strongly constrained since residual annihilation after freeze-out must not disrupt the consistency of light element abundance predictions from big bang nucleosynthesis (BBN), precision observations of the cosmic microwave background (CMB), and X/$\gamma$-rays in the Milky Way. Indeed, consistency with the observed CMB anisotropies rules out thermal relic models with a mass below $\sim 5$~GeV~\cite{Slatyer:2009yq,Madhavacheril:2013cna} unless $\langle \si v\rangle$ is suppressed in that epoch, for example by the low characteristic kinetic energy. Thus, sub-GeV DM models generically require dominant $p$-wave annihilation channels or mass thresholds which similarly limit annihilation at low-velocities. Late annihilation in the halo of the Milky Way or satellite dwarf spheroidal galaxies is also constrained by indirect detection searches, although the sub-GeV regime is subject to somewhat weaker constraints due to the `MeV gap' in observations below the sensitivity threshold of the Fermi satellite~\cite{Gonzalez-Morales:2017jkx}.

The attention focused on sub-GeV dark matter models with suppressed low velocity annihilation in turn heightens the potential significance of BBN as the regime with perhaps the largest characteristic temperature. Thus, in this paper we re-examine BBN sensitivity to sub-GeV dark matter annihilation, focusing on models with dominant $p$-wave channels. BBN has been studied extensively as a probe of new physics scenarios with late decays of long-lived particles~\cite{Sarkar:1995dd,Schramm:1997vs,Jedamzik:2006xz,Kawasaki:2008qe,Iocco:2008va,Pospelov:2010hj,Fradette:2017sdd,Kawasaki:2017bqm,Ghosh:2022frt,Dev:2025pru}, while its utility as a probe of annihilation has received less attention as it is generally less sensitive than other observables, particularly for higher-mass DM candidates~\cite{Jedamzik:2004er,Jedamzik:2009uy,Henning:2012rm,Depta:2019lbe}. However, BBN does provide quite a robust limit on the lower mass threshold for thermal relic dark matter, with recent estimates putting this at the scale of 3--10~MeV depending on annihilation channels~\cite{Sabti:2019mhn,Chu:2022xuh,Chu:2023jyb}, because freeze-out after neutrino decoupling is highly constrained. Here we focus on a slightly higher sub-GeV mass range, between $\sim100$--$1000$~MeV, where DM freeze-out occurs before the stages of neutrino decoupling and electron-positron annihilation. We find that the relatively long-lived products of residual hadronic annihilation, charged pions and kaons, can still impact the pre-bottleneck era of BBN at a non-negligible level. While the current observational precision on the D and $^4$He abundances is not quite at the level required to impose constraints on benchmark models of thermal relic sub-GeV dark matter with $p$-wave annihilation, it is superior to other indirect channels and provides comparable sensitivity to a number of terrestrial probes. Thus, it is an interesting complementary tool and the sensitivity can be enhanced with higher precision data on light element abundances.

The rest of this paper is structured as follows. Section~2 reviews the impact of hadronic injection on BBN, focusing on charged pseudoscalar pions and kaons injected either before or after the deuterium bottleneck. In the more relevant pre-bottleneck regime, we track the impact of pion and kaon injection from dark matter annihilation on D and $^4$He abundances due to proton/neutron charge conversion. In Section~3, we compare the impact of dark matter annihilation in hadronic channels to the current observational limits, and then apply these results to one of the benchmark sub-GeV dark matter models: a scalar candidate with dark photon mediation that exhibits freeze-out through $p$-wave annihilation. The results provide interesting complementarity to direct terrestrial probes. We finish with some concluding remarks in Section~4.

\vspace{-2mm}
\section{Hadronic injection into BBN}

For thermal relic scenarios, dark matter annihilation has negligible impact on the thermal history of the Universe, but the injection of hadrons into the thermal medium during BBN can have a significant impact on the nuclear reaction network and the light element abundances. More specifically, charged mesons have the largest early effects on BBN outcomes through their direct interactions with nucleons, and charge conversion reactions in particular~\cite{Sarkar:1995dd,Schramm:1997vs,Jedamzik:2006xz,Kawasaki:2008qe,Iocco:2008va,Pospelov:2010hj,Pospelov:2010cw,Kawasaki:2017bqm}. In this section, we consider the two lightest charged mesons, $\pi^\pm$ and $K^\pm$, and assume that dark matter annihilates entirely into one species or the other, with an overall relic density fixed to the observed thermal value, $\Omega_{\rm DM} h^2 \simeq 0.12$. Accounting for branching ratios into specific hadronic states can subsequently facilitate mapping the results onto more complete models, but this setup captures the leading modifications to BBN and allows us to assess the generic sensitivity to the DM annihilation rate. 

To perform BBN calculations, we adapt the numerical framework developed in Ref.~\cite{McKeen:2024voa}, which agrees well with other well established codes~\cite{Pitrou:2019nub,Burns:2023sgx}. The nuclear reaction network we solve includes the light species most relevant to BBN—namely, neutrons, protons, deuterium, tritium, ${}^3$He, and ${}^4$He. Since our analysis focuses on the abundances of deuterium and ${}^4$He, which provide the most precise observational constraints, we restrict the network to these light nuclides. The inclusion of heavier elements like ${}^7$Li and ${}^7$Be, which are subject to potential ambiguities such as the Lithium problem~\cite{Molaro:2024wxa,Ryan:1999rxn,Olive:1999ij}, provides only subleading sensitivity to the early effects of charge conversion and will not be explored in detail here. 

We impose the assumptions of standard BBN (SBBN) \cite{McKeen:2024voa,RevModPhys.88.015004,Kolb:1990vq,Copi:1994ev}. These include a radiation-dominated $\Lambda$ cold dark matter ($\Lambda$CDM) expansion history; three neutrino species; and input parameters fixed by independent measurements: the baryon-to-photon ratio $\eta_b \simeq 6.1\times 10^{-10}$~\cite{Planck:2018vyg, Fields:2019pfx}, the neutron lifetime $\tau_n \simeq 878.4$ s~\cite{ParticleDataGroup:2024cfk}, and various nuclear reaction and weak rates for the processes involved~\cite{Serpico:2004gx,Nagai:2006jb}. We further assume that neutrinos decouple instantaneously from the plasma prior to electron-positron annihilation. Small corrections due to non-instantaneous neutrino decoupling are a subleading effect at the precision level of this work. Nevertheless, we approximate those effects by setting the number of neutrino species to $N_\nu=3.045$, to produce an effective number of relativistic species $N_{\rm eff}$ consistent with non-instantaneous neutrino decoupling treatments~\cite{deSalas:2016ztq,Escudero:2018mvt}. With these assumptions, the code reproduces standard BBN predictions for deuterium and helium abundances within the expected uncertainties, which we use as a baseline to assess the impact of dark matter annihilation.

The impact of both pion and kaon injection on BBN will be examined in what follows, but initially we focus solely on charged pion injection. We therefore assume that DM annihilation and freeze-out proceed via a single channel, $\ch\ch \leftrightarrow \pi^+\pi^-$. Pions remain in equilibrium with the thermal plasma well after DM freeze-out, maintained for example at high temperatures by $\pi^+\pi^-$ annihilation processes. The equilibrium abundance then drops rapidly as the temperature cools below $T \sim m_\pi \simeq 140$~MeV and pions become non-relativistic. However, once the pion abundance drops below that of nucleons, charge exchange reactions with nucleons become more significant. This is the point at which we will start our analysis, solving for the subsequent evolution of the pion number density $n_\pi$ after both DM freeze-out and the point at which the pion abundance drops well below that of nucleons. We utilize Boltzmann equations (one for $\pi^+$ and one for $\pi^-$) that are coupled to the BBN network. Schematically, the pion Boltzmann equation is of the form,
\begin{align}
    \begin{aligned}
	\frac{dn_\pi}{dt} + 3H n_\pi =\ & \rm DM\ injection\ -\ pion\ decay\ \\
    & + \rm pion/nucleon\ int.\ + \cdots
    \end{aligned},
    \label{eqn}
\end{align}
and we now discuss the terms on the right-hand side in more detail. 

The contribution from residual dark matter injection through post-freeze-out annihilation $\ch\ch \rightarrow \pi^+\pi^-$ is simply $\langle \sigma v \rangle_{\rm ann} n_\chi^2$, where the dark matter number density is given by 
\begin{equation}
	n_\chi = n_{\chi,0}\left( \frac{T}{T_0}\right)^3 \simeq \left[ 0.12 \frac{\rho_{\rm crit,0}}{m_\chi} \right] \left( \frac{T}{T_0}\right)^3,
\end{equation}
and quantities with a $0$ subscript are measured today. In practice, it is convenient to normalize all number densities to the density of baryons,
\be
 X_i \equiv \frac{n_i}{n_b}.
\ee
Following the BBN formalism of \cite{McKeen:2024voa}, the term in the Boltzmann equation due to injection from DM annihilation then takes the form,
\begin{equation}
	\left .\dot{X}_{\pi^\pm}\right|_{\rm DM\, inj} = \Gamma_{\chi}\frac{X_\chi^2}{2},
\end{equation}
where $\dot{X} \equiv dX/dt$. For $p$-wave dark matter annihilation, as required to avoid CMB constraints, we have 
\begin{equation}
	\Gamma_{\chi} \equiv \langle \sigma v \rangle_{\rm ann} n_b = \frac{3}{2} \frac{T}{m_\ch} b \  \eta_b n_\gamma,
\end{equation}
where $n_\gamma$ is the photon number density, and $b$ characterizes the size of the $p$-wave annihilation rate,
\be
\langle \sigma v \rangle_{\rm ann}^{p-{\rm wave}} = \frac{3}{2} \frac{T}{m_\ch} b, 
\ee
and thus the injection rate of pions into the pre-BBN medium.

Pions freely decay with a lifetime of $2.6 \times 10^{-8}$ s~\cite{ParticleDataGroup:2024cfk}, and so the pion decay term in the Boltzmann equation is simply,
\begin{align}
   \left . \dot{X}_{\pi^\pm}\right|_{\rm decay} = \frac{X_{\pi^\pm}}{\tau_{\pi^\pm}}. 
\end{align}
In principle, the pion lifetime should be adjusted to account for relativistic effects for in-flight pions $\tau = \tau_{\pi^\pm} \gamma(E_\pi (t))$, however the assumption of stopped pions holds strongly in this case, as discussed below. 

The final term on the right-hand side of (\ref{eqn}) accounts for charge conversion with nucleons, and thus couples the evolution of pions to the BBN network. Charged pions induce the following charge conversion reactions~\cite{Pospelov:2010cw}:
\begin{align}
    \begin{aligned}
	\pi^- + p \to n + \gamma, \quad 	& (\sigma v)^{\pi^-}_{pn\gamma} \simeq 0.57\ \rm mb, \\
	\pi^- + p \to n + \pi^0 ,\quad	& (\sigma v)^{\pi^-}_{pn\pi^0} \simeq 0.88\ \rm mb, \\
	\pi^+ + n \to p + \pi^0, \quad 		& (\sigma v)^{\pi^+}_{np\pi^0} \simeq 1.7\ \rm mb. 	
\end{aligned}
\end{align}
The first two reactions experience Coulomb enhancement, so that the thermal cross section for those processes is given by
\begin{equation}
\begin{aligned}
    (\sigma v)_{\rm th}^{\pi^-} & \equiv F_{p\pi^-} \Big(
	(\sigma v)^{\pi^-}_{pn\gamma} + (\sigma v)^{\pi^-}_{pn\pi^0}
	\Big) \\
    & \simeq F_{p\pi^-} \times 1.45\ \rm mb,
\end{aligned}
\end{equation}
where the Coulomb correction factor $F_{p\pi^-}\simeq 2$ in the temperature regime of interest for BBN~\cite{Pospelov:2010cw}. In each of the reactions above, we will ignore backward reactions, as they are generally much weaker. These contributions to the Boltzmann equations are then summarized as follows,
\begin{align}
	\begin{aligned}
		\dot{X}_{p\pi^-} & = \Gamma_{p\pi^-} X_{\pi^-} X_p, \quad && \Gamma_{p\pi^-} \equiv (\sigma v)_{\rm th}^{\pi^-} n_b, \\
		\dot{X}_{n\pi^+} & = \Gamma_{n\pi^+} X_{\pi^+} X_{n}, \quad &&  \Gamma_{n\pi^+} \equiv  (\sigma v)^{\pi^+}_{np\pi^0} n_b.
	\end{aligned}
\end{align}

\begin{figure}[t]
    \centering
\includegraphics[width=0.99\linewidth]{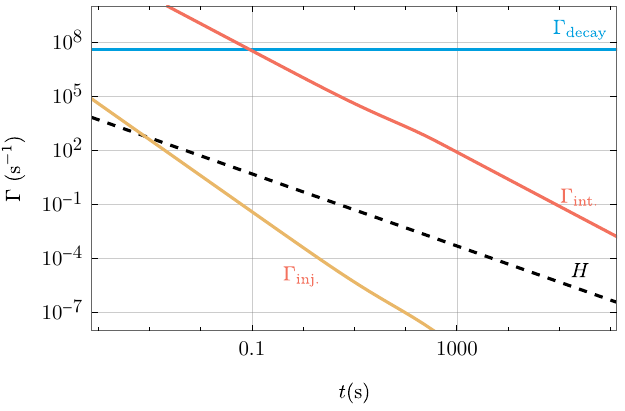}
    \caption{Comparison of the rates that go into the pion Boltzmann equation, showing the pion decay rate (blue), pion-nucleon interaction rate (red), and the pion injection rate due to DM annihilation (yellow) with $m_\chi = 400$ MeV, and $b=10^{-23}\ \rm cm^3/s$. The Hubble parameter (black, dashed) is also shown for reference.}
    \label{Rates}
\end{figure}

Putting all the components together, the Boltzmann equations (\ref{eqn}) take the form
\begin{align}
	\dot{X}_{\pi^-}	&= \Gamma_{\chi \pi} \frac{X_\chi^2}{2} - \frac{X_{\pi^-}}{\tau_{\pi^\pm}} - \dot{X}_{p\pi^-}, \\
	\dot{X}_{\pi^+}	&= \Gamma_{\chi \pi} \frac{X_\chi^2}{2} - \frac{X_{\pi^+}}{\tau_{\pi^\pm}} - \dot{X}_{n\pi^+},
\end{align}
and the various rates are illustrated as a function of time/temperature in Fig.~\ref{Rates}.
Corresponding terms must also be added to the neutron and proton equations for consistency:
\begin{align}
	\dot{X}_p & \supset - \dot{X}_{p\pi^- }  +  \dot{X}_{n\pi^+}, \\
	\dot{X}_n & \supset + \dot{X}_{p\pi^-} -  \dot{X}_{n\pi^+}.
\end{align}
This set of equations describes the reaction network to be analyzed below. We note that further processes could be included on the right-hand side of (\ref{eqn}), for example $\pi^+\pi^-$ annihilation to Standard Model states, which plays a leading role at high temperature but which is subleading well after the pion abundance drops below that of nucleons. We comment further on various subleading processes at the end of this sub-section.

\begin{figure}[t]
    \centering
\includegraphics[width=0.99\linewidth]{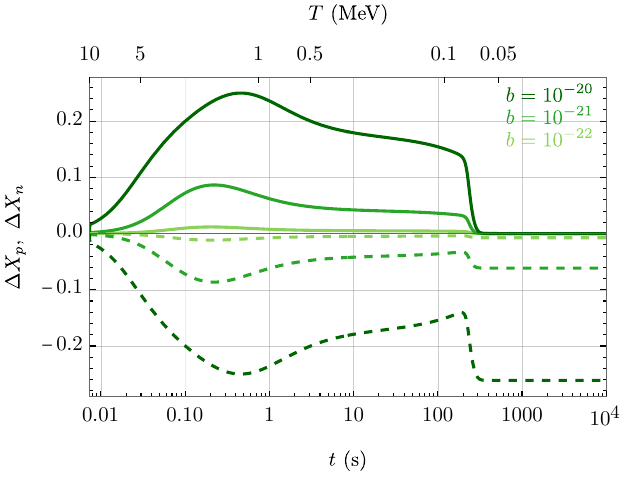}
    \caption{Evolution in the difference $\Delta X = X - X|_{\rm SBBN}$ in the fractions of neutrons (solid) and protons (dashed) away from the SBBN values when pions are injected with DM mass $m_\chi=400\ \rm MeV$ and varying DM annihilation rate $b$.  } 
    \label{fig:Deltas}
\end{figure}

The effects of injected pions on neutron (solid) and proton (dashed) abundances are shown in \cref{fig:Deltas}, for a fixed DM mass $m_\chi = 400$ MeV, and varying DM annihilation rate $b$. Larger DM annihilation rates (darker curves) result in larger differences due to the increased number of injected pions. Due to the Coulomb correction factor $F_{\textrm{p}\pi^-}$, charge conversion reactions bias the conversion from protons to neutrons, thus increasing the number of available neutrons compared to SBBN before the bottleneck. A mirrored decrease in the proton fraction occurs since no other nuclei have been sufficiently produced at this point. The differences reach a peak then reverse direction as the injected pions are depleted and regular weak rates become the dominant interaction between protons and neutrons again, but with a larger neutron fraction than SBBN, thus biasing proton production processes and resulting in a decrease in $X_n$, and corresponding increase in $X_p$. The peaks are reached at later times for larger DM annihilation rates $b$, since larger rates correspond to larger pion abundance, and correspondingly a longer period is required to deplete enough pions before weak rates become dominant. 

\begin{figure}[t]
    \centering
    \includegraphics[width=0.99\linewidth]{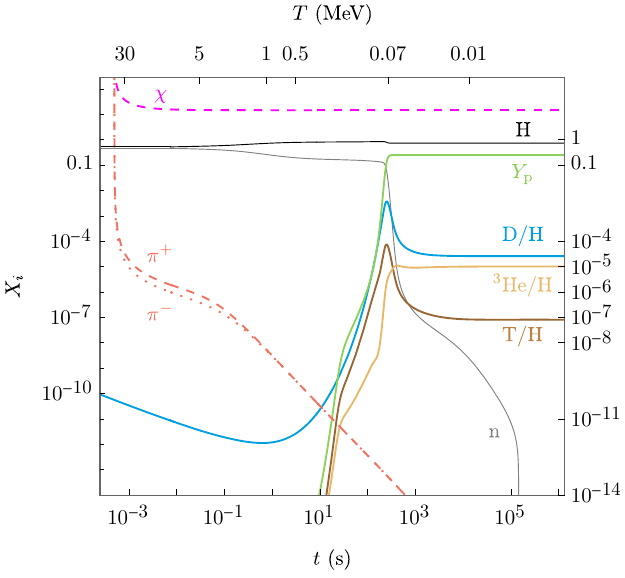}
    \caption{The evolution of abundances with DM freeze-out governed by annihilation to pions with $m_\chi = 400$ MeV, and $b = 10^{-23}\ \rm cm^3/s$. The evolution of the pion number densities is described in the text.}
    \label{fig:DMExplicit}
\end{figure}

The pion channel is only possible above the kinematic threshold, $m_\chi \geq m_\pi$. In this mass range, thermal dark matter freeze-out occurs at a temperature $T_f \gtrsim 5$ MeV, before significant BBN processes begin. The evolution of the abundances of the DM (magenta), pions (red), and various nuclear species produced due to BBN are shown in \cref{fig:DMExplicit} for a choice of $b= 10^{-23}\ \rm cm^3/s$ and $m_\chi = 400$ MeV. Due to the different pion-nucleon interactions that positively and negatively charged pions undergo we see some deviation in their evolution at early times when interactions with nucleons are dominant over decays. The dominant interaction rates for charged pions at very early times (from DM injection), and at later times (pion decay) are independent of charge, and therefore their abundances become very similar in those periods. The pions begin their evolution in chemical equilibrium with DM at high temperatures before DM freeze-out. At around $t\sim 10^{-3}$ s, they decouple from DM and their evolution becomes dominated by charge conversion interactions up to $t\sim 0.1$ s, when pion decay becomes the dominant rate as shown in \cref{Rates}. Pion decay rate equilibrates with the residual injection from DM which explains the power law decay at later times, rather than an exponential as in the case of neutrons for example. We note here that DM freeze-out is always fixed to yield the correct relic abundance at late times.

The same analysis can be applied to DM annihilating to kaons. Charged kaons (with mass $m_{K^\pm} = 493.7\ {\rm MeV}$ and lifetime $\tau_{K^\pm} = 1.24\times 10^{-8}\ \rm s$ \cite{ParticleDataGroup:2024cfk}) also contribute to charge conversion reactions, and the charge exchange cross sections for kaons are even higher than for pions \cite{Pospelov:2010cw,Reno:1987qw}. However, due to their larger mass, the annihilation channel is only accessible at higher dark matter masses. Charged kaons have similar decay rates to pions, so we expect the decay to dominate over DM injection, and possibly compete with kaon-nucleon interactions at earlier times. Further, since only larger DM masses are viable, kaons are characteristically injected at earlier times where they are rapidly stopped; indeed, kaons are efficiently stopped at slightly higher temperatures than pions (see, e.g., Fig. 1 of Ref~\cite{Pospelov:2010cw}). Although we will not study them here, neutral kaons have longer lifetimes (for $K_L$), but their effect on BBN is similar to charged kaons~\cite{Pospelov:2010cw}. 

The most significant impact of charged kaons on BBN is through their charge conversion reactions. The reaction $K^- + p \to \bar{K}^0+n$ is not kinematically allowed for stopped kaons, and so it is heavily suppressed, but generally charge conversion reactions can proceed via exchange reactions with hyperons in the final state that decay to pions and nucleons. Effectively, the kaon's charge conversion rate is~\cite{Pospelov:2010cw}
\begin{align}
\begin{aligned}
    	& K^- + p \to n + X:\quad (\sigma v)_{pn}^{K^-} \simeq 32\ \rm mb,
	\label{KCoul}
	\\
	& K^- + n \to p + X:\quad (\sigma v)_{np}^{K^-} \simeq 13\ \rm mb.
    \end{aligned}
\end{align}
The first reaction in (\ref{KCoul}) also receives a contribution from Coulomb corrections. Notably, the nucleons in the final state can have energies on the order of 30 MeV. While the protons will be rapidly stopped before inducing any changes to BBN, the neutrons could in principle affect other processes, e.g. via hadro-dissociation after the bottleneck. However, such effects are generally negligible since most neutrons would be stopped through interacting with protons. 

It is also important to account for kaon decay to charged pions. Kaons decay to pions with branching ratios~\cite{ParticleDataGroup:2024cfk},
\begin{align}
\begin{aligned}
        K^\pm \to \pi^\pm \pi^0 (\pi^0) \quad & {\rm BR} =22.4\%, \\
    K^\pm \to
     \pi^\pm \pi^\pm \pi^\mp \quad &  {\rm BR} =5.6\%. 
     \end{aligned}
\end{align}
These channels provide an additional source term in the Boltzmann equation for pions,
\begin{equation}
    \dot{X}_{\pi^\pm} \supset {\rm BR_{eff}} \frac{X_{K^\pm}}{\tau_{K^\pm}},
\end{equation}
where $\rm{BR_{eff}} = 0.336$ is the effective branching ratio from both decay modes after accounting for multiplicities.

Overall, we find that kaons have a stronger impact on BBN than pions, but injection is only viable for dark matter masses above the kaon kinematic threshold ($m_{\chi} \geq m_K$). We will account for this channel in considering the sensitivity to DM annihilation in the next section.

For completeness, we now review several other processes which, while often relevant for BBN overall, were found to be sub-leading for the early charge exchange processes that are most significant here:

\begin{itemize}
\item For dark matter with mass $m_\chi \gtrsim m_\pi$, freezout occurs before neutrino decoupling and therefore well before the deuterium bottleneck. As a result, the pion abundance has dropped substantially by the time the light elements start to populate with significant abundances (see \cref{fig:DMExplicit}). Although still present, meson interactions with other nuclei thus have a negligible effect on BBN predictions. Those effects are nonetheless included in the full BBN network for completeness. 

\item Depending on the DM mass, the injected mesons could in principle have a non-negligible amount of kinetic energy. This can significantly impact their interaction cross section with nucleons and various nuclei, e.g. through excitation of $\Delta$ resonances. For the present scenario, however, meson injection is most important well before the bottleneck. At these temperatures, with $T\gtrsim m_e$, Coulomb interactions are highly efficient at stopping mesons, and the relevant hierarchy of rates satisfies $\Gamma_{\rm stop} \gg \Gamma_{\rm dec} \gg \Gamma_{\rm p,n} \gg H$. This hierarchy is illustrated in \cref{Rates} for $m_\chi = 400$ MeV, and $b=10^{-23}\ \rm cm^3/s$. Pion stopping is not shown in \cref{Rates} as in this scenario it scales as $\Gamma^\pi_{\rm stop} \sim 10^{15}\ {\rm s^{-1}} (T/{1\ \rm MeV})^2$, and thus is dominant.

\item We neglect final-state radiation (FSR) in pion and kaon decays, which in principle could result in significant electromagnetic energy injection into the BBN network. However, as most mesons decay well before any nuclei form, the effects of photo-dissociation processes are suppressed. Thermal effects from radiation onto the plasma or from other annihilation products are also negligible due to the small baryon-to-photon ratio. A full treatment with non-instantaneous neutrino decoupling could be sensitive to FSR. However, since mesons are very efficiently stopped, and therefore their short lifetimes are not extended by a time dilation factor, the majority of injected mesons decay well before the neutrino decoupling phase for $m_\chi > m_\pi$, thus limiting potential impacts of FSR. 
\end{itemize}

\vspace{-2mm}
\section{Limits on DM annihilation}
\label{sec:Generic Constraints}

 Since the most significant effect pions have on BBN occurs around $T\sim$ MeV, it is helpful to qualitatively estimate the pion abundance required during this period to noticeably affect BBN outcomes. At $t \sim 0.1$~s, proton--neutron conversion via weak interactions occurs at a rate roughly $\sim G_F^2 T^5 \sim 10\ \rm s^{-1}$. The pion--nucleon interaction rate at the same time is $\Gamma_{p\pi} \sim 10^7\ \rm s^{-1}$, so for these processes to compete with the standard weak rates, the pion fraction must be of order $\mathcal{O}( 10^{-6})$, consistent with the pion abundance at that time shown in \cref{fig:DMExplicit}. Given that SBBN predictions agree at the $\sim$1\% level with the observed abundances of light nuclei, this estimate hints that residual DM injection may produce measurable effects. In this section, we quantify the sensitivity first in a general case and then for a specific sub-GeV model.

\vspace{-2mm}
\subsection{Generic constraints on $p$-wave annihilating DM}

\begin{figure}[t]
    \centering
    \includegraphics[width= 0.99\linewidth]{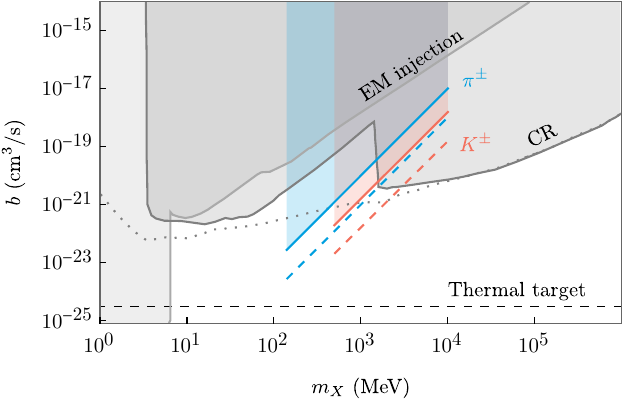}
    \caption{The $95\%$ CL sensitivity to $p$-wave DM annihilation to pions (blue) and kaons (red) during BBN. Note that we set the DM freeze-out abundance to its observed value, and then the region above the contours is nominally excluded by requiring consistency with BBN's prediction of D/H and $Y_p$ compared to observations~\cite{PDG2020,Cooke:2013cba,*Riemer-Sorensen:2014aoa,*Balashev:2015hoe,*Riemer-Sorensen:2017pey,*2018MNRAS,*Cooke:2017cwo, Aver:2020fon,*Valerdi:2019beb,*Fernandez:2019hds,*Kurichin:2021ppm,*2020ApJ,*2021MNRAS,*2022MNRAS}. Dashed colored curves indicate a branching ratio of 1, while solid curves indicate a branching ratio of 0.1 with the remainder of annihilation products neglected. Also shown are limits from EM injection to BBN~\cite{Depta:2019lbe}, and cosmic ray injection of positrons from Voyager 1 and AMS-02 data (with the dotted curve indicating a less conservative cosmic ray propagation model)~\cite{Boudaud:2018oya}. The gray-dashed horizontal line indicates the value of $b$ required for thermal freeze-out to match the measured DM abundance.}
    \label{fig:blimit}
\end{figure}

Without specifying a model, making a choice for DM mass $m_\chi$ and interaction strength $b$ fixes the the DM--meson interaction rate $\Gamma_\chi$. Although in a thermal freeze-out picture this also fixes the DM freeze-out abundance, we instead fix DM abundance post-freeze-out to correspond to the measured DM relic abundance that yields $\Omega_{\rm DM} h^2 \simeq 0.12$. Since meson injection due to DM annihilation is the only source term for the mesons, this choice consequently fixes all new inputs into BBN. For each choice in the parameter space, we compare the predicted abundances of deuterium and $^4$He with the observed values of of the deuterium fraction, ${\rm D/H} \times 10^{-5}=  2.547 \pm 0.025$, and $^4$He mass fraction, $Y_p = 0.245\pm 0.003$~\cite{PDG2020,Cooke:2013cba,*Riemer-Sorensen:2014aoa,*Balashev:2015hoe,*Riemer-Sorensen:2017pey,*2018MNRAS,*Cooke:2017cwo, Aver:2020fon,*Valerdi:2019beb,*Fernandez:2019hds,*Kurichin:2021ppm,*2020ApJ,*2021MNRAS,*2022MNRAS}. Combining both observations, we calculate the $2\sigma$ limits on the interaction strength $b$ for the mass range $m_{\pi,K}\leq m_\chi \leq 10$ GeV, as shown in \cref{fig:blimit}. The lines show upper limits on the cross section parameter $b$, where blue lines are limits from annihilations to pions, and red lines are limits from annihilations to kaons (with subsequent kaon decay to pions accounted for). The dashed lines indicate annihilations with BR~$=1$, while solid lines indicate BR~$=0.1$, where the rest of the annihilation products are assumed to not have a significant effect on BBN predictions. Leading limits from electromagnetic injection into BBN~\cite{Depta:2019lbe} and limits from cosmic ray injection from Voyager 1 and AMS-02~\cite{Boudaud:2018oya} are also shown in gray regions. On the lower mass end, the limits cut off at the kinematic thresholds $m_{\chi} = m_{\pi^\pm}$ and $m_\chi = m_{K^\pm}$ for pions and kaons respectively, and on the high mass end, the limits extend to $m_\chi = 10$ GeV for illustrative purposes, although at masses $m_\chi \gtrsim 1$ GeV these fixed branching ratios are less realistic due to the opening up of many hadronic decay channels.  

\vspace{-2mm}
\subsection{Limits on a DM benchmark model}

As a benchmark, we consider DM annihilation to SM final states, mediated by a dark photon $A'$ which mixes with the photon, thus coupling it to most SM particles. A hidden $U(1)_d$ gauge field $A'_\mu$ can kinetically mix with the SM hypercharge $B_{\mu\nu}$ via the term \cite{Holdom:1985ag}
\begin{equation}
    \mathcal{L}\supset \frac{\epsilon}{2\cos\theta_W}B_{\mu\nu}F'^{\mu\nu},
\end{equation}
where $\epsilon$ is the kinetic mixing parameter, $F'$ is the field strength tensor for the dark photon, and $\theta_W$ is the weak mixing angle. 
After symmetry breaking, including a current for the DM particle gives the interaction Lagrangian,
\begin{equation}
    \mathcal{L} = -A'_\mu\left( g_D J_D^\mu + \epsilon e J_{\rm EM}^\mu \right),
\end{equation}
where $g_D$ is the DM-$A'$ coupling constant, and $J^\mu_{D,\rm EM}$ are the dark and electromagnetic currents, respectively.
Since this work focuses on $p$-wave annihilations, we consider a model with a scalar dark matter candidate $\ch$, where
\begin{equation}
    J_D^\mu = i\chi^* \partial^\mu \chi + c.c.
\end{equation}

For the relevant dark matter mass range, and for sufficiently small kinetic mixing, the annihilation cross section can be approximated as \cite{deNiverville:2012ij}
\begin{equation}
    \langle \sigma v \rangle_{\rm ann} \simeq \langle \sigma v \rangle_e + \langle \sigma v \rangle_\mu +\left\langle (\sigma v)_\mu R(s=4m_\chi^2)\right\rangle,
    \label{DMCS}
\end{equation}
where $\langle \si v\rangle_{e,\mu} \propto \ep^2 \al \al_D$, with $\alpha_D = g_D^2/4\pi$, is the dark photon mediated annihilation rate into leptons \cite{Boehm:2003hm}, and $R \equiv \sigma_{e^+e^-\to \rm hadrons}/\sigma_{e^+e^-\to \mu^+\mu^-}$ accounts for hadronic annihilation channels.

Only a fraction of the annihilated DM has hadronic final states. The injection of an annihilation product $j$ into the plasma is governed by a term in the Boltzmann equation,
\begin{equation}
    \Gamma_j = \langle \sigma v \rangle_{\rm ann} \textrm{BR}_{\rm eff}^j ~n_b,
\end{equation}
where $\textrm{BR}_{\rm eff}^j$ is the (multiplicity-weighted) effective branching ratio for DM annihilation modes with $j$ in the final state. The annihilations we consider mostly proceed off-shell, but for simplicity we approximate this branching ratio to be the same as the branching ratio for a dark photon decay to various final states \cite{Batell:2009yf}.

\begin{figure}[t]
    \centering
    \includegraphics[width=0.99\linewidth]{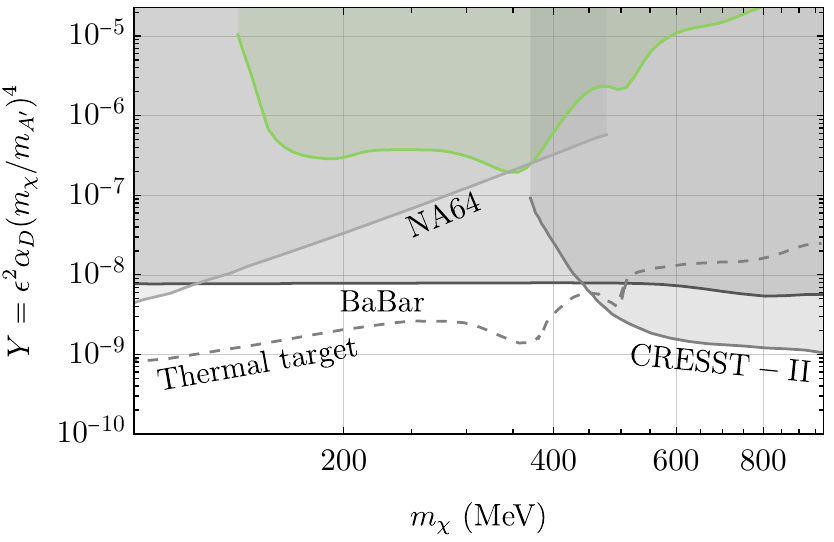}
    \caption{The $90\%$ CL limits on DM annihilation to pseudoscalar mesons during BBN (green). As in Fig.~\ref{fig:blimit}, we set the DM freeze-out abundance to its observed value, and then the region above the contour is  nominally excluded by BBN's prediction of D/H and $Y_p$ compared to observations~\cite{PDG2020,Cooke:2013cba,*Riemer-Sorensen:2014aoa,*Balashev:2015hoe,*Riemer-Sorensen:2017pey,*2018MNRAS,*Cooke:2017cwo, Aver:2020fon,*Valerdi:2019beb,*Fernandez:2019hds,*Kurichin:2021ppm,*2020ApJ,*2021MNRAS,*2022MNRAS}. Also shown are limits from BaBar~\cite{BaBar:2017tiz}, NA64~\cite{NA64:2023wbi}, and CRESST-II~\cite{CRESST:2015txj} (gray regions), and the thermal relic line (gray-dashed curve) where the choice of parameters corresponds to the observed DM relic abundance~\cite{Berlin:2018bsc}. A conventional slice of the dark photon model parameter space is shown with the $\alpha_D=0.5$ and $m_{A'}=3m_\chi$ fixed.}
    \label{fig:DarkPhotonLimits}
\end{figure}

An appropriate choice of the four parameters in this model fixes the rate of injection $\Gamma_j$. The treatment hereafter is the same as in~\cref{sec:Generic Constraints}, where the effects of hadronic injection propagate throughout the BBN nuclear reaction network as we solve the set of coupled Boltzmann equations. For the benchmark choice of $\alpha_D = 0.5$, and $m_{A'}/m_\chi = 3$, the limits on DM kinetic mixing are shown by the green contour in \cref{fig:DarkPhotonLimits}, where the shaded region above it is excluded by consistency with the BBN predictions for D/H and $Y_p$ abundances.

\section{Discussion}

In this paper, we have revisited the hadronic annihilation channels of thermal relic dark matter, focusing on the sub-GeV mass range for which $p$-wave annihilation is relevant due to CMB constraints. We found that the combination of relatively late DM freeze-out with the long lifetime of charged pions and kaons allows these hadronic annihilation products to play a significant role impacting neutron-proton conversion in the early pre-bottleneck phase of BBN. Although the current observational sensitivity to the abundance of D and $^4$He is not sufficient to fully constrain the thermal relic annihilation rate, it is significantly greater than other indirect probes such as the CMB and galactic annihilation. Indeed, for the benchmark models of dark-photon-mediated sub-GeV dark matter, this BBN sensitivity channel is quite complementary to that of collider and direct detection experiments. 

While the requirement of $p$-wave annihilation is one generic means of protecting sub-GeV thermal relics from stringent CMB constraints, another is to require a small mass splitting $\De m$ between the annihilating components of e.g. pseudo-Dirac dark matter. As the dominant impact of hadronic injection occurs early in the pre-bottleneck phase, BBN can impose important constraints on $\De m$ if the $s$-wave annihilation rate is sizable. We observe from Fig.~\ref{fig:Deltas} that $\De m$ should be larger than the characteristic kinetic energy at temperatures up to $T \sim$ 2--3~MeV, which notably lies above the electron-positron threshold. It would also be interesting to explore the consequences for inelastic DM candidates and other sub-GeV thermal relic candidates with mediators beyond the dark photon. For example, BBN may provide interesting sensitivity to sub-GeV DM models with vector mediators such as gauged $B-3L_\tau$, which are subject to less stringent constraints from terrestrial searches.

\vspace{-2mm}

\section{Acknowledgments}

We are grateful to David McKeen and Maxim Pospelov for helpful discussions and comments on the manuscript. This work was supported in part by NSERC, Canada.

\bibliography{References}

\begin{thebibliography}{62}%
\makeatletter
\providecommand \@ifxundefined [1]{%
 \@ifx{#1\undefined}
}%
\providecommand \@ifnum [1]{%
 \ifnum #1\expandafter \@firstoftwo
 \else \expandafter \@secondoftwo
 \fi
}%
\providecommand \@ifx [1]{%
 \ifx #1\expandafter \@firstoftwo
 \else \expandafter \@secondoftwo
 \fi
}%
\providecommand \natexlab [1]{#1}%
\providecommand \enquote  [1]{``#1''}%
\providecommand \bibnamefont  [1]{#1}%
\providecommand \bibfnamefont [1]{#1}%
\providecommand \citenamefont [1]{#1}%
\providecommand \href@noop [0]{\@secondoftwo}%
\providecommand \href [0]{\begingroup \@sanitize@url \@href}%
\providecommand \@href[1]{\@@startlink{#1}\@@href}%
\providecommand \@@href[1]{\endgroup#1\@@endlink}%
\providecommand \@sanitize@url [0]{\catcode `\\12\catcode `\$12\catcode `\&12\catcode `\#12\catcode `\^12\catcode `\_12\catcode `\%12\relax}%
\providecommand \@@startlink[1]{}%
\providecommand \@@endlink[0]{}%
\providecommand \url  [0]{\begingroup\@sanitize@url \@url }%
\providecommand \@url [1]{\endgroup\@href {#1}{\urlprefix }}%
\providecommand \urlprefix  [0]{URL }%
\providecommand \Eprint [0]{\href }%
\providecommand \doibase [0]{https://doi.org/}%
\providecommand \selectlanguage [0]{\@gobble}%
\providecommand \bibinfo  [0]{\@secondoftwo}%
\providecommand \bibfield  [0]{\@secondoftwo}%
\providecommand \translation [1]{[#1]}%
\providecommand \BibitemOpen [0]{}%
\providecommand \bibitemStop [0]{}%
\providecommand \bibitemNoStop [0]{.\EOS\space}%
\providecommand \EOS [0]{\spacefactor3000\relax}%
\providecommand \BibitemShut  [1]{\csname bibitem#1\endcsname}%
\let\auto@bib@innerbib\@empty
%</preamble>
\bibitem [{\citenamefont {Lee}\ and\ \citenamefont {Weinberg}(1977)}]{Lee:1977ua}%
  \BibitemOpen
  \bibfield  {author} {\bibinfo {author} {\bibfnamefont {B.~W.}\ \bibnamefont {Lee}}\ and\ \bibinfo {author} {\bibfnamefont {S.}~\bibnamefont {Weinberg}},\ }\bibfield  {title} {\bibinfo {title} {{Cosmological Lower Bound on Heavy Neutrino Masses}},\ }\href {https://doi.org/10.1103/PhysRevLett.39.165} {\bibfield  {journal} {\bibinfo  {journal} {Phys. Rev. Lett.}\ }\textbf {\bibinfo {volume} {39}},\ \bibinfo {pages} {165} (\bibinfo {year} {1977})}\BibitemShut {NoStop}%
\bibitem [{\citenamefont {Slatyer}\ \emph {et~al.}(2009)\citenamefont {Slatyer}, \citenamefont {Padmanabhan},\ and\ \citenamefont {Finkbeiner}}]{Slatyer:2009yq}%
  \BibitemOpen
  \bibfield  {author} {\bibinfo {author} {\bibfnamefont {T.~R.}\ \bibnamefont {Slatyer}}, \bibinfo {author} {\bibfnamefont {N.}~\bibnamefont {Padmanabhan}},\ and\ \bibinfo {author} {\bibfnamefont {D.~P.}\ \bibnamefont {Finkbeiner}},\ }\bibfield  {title} {\bibinfo {title} {{CMB Constraints on WIMP Annihilation: Energy Absorption During the Recombination Epoch}},\ }\href {https://doi.org/10.1103/PhysRevD.80.043526} {\bibfield  {journal} {\bibinfo  {journal} {Phys. Rev. D}\ }\textbf {\bibinfo {volume} {80}},\ \bibinfo {pages} {043526} (\bibinfo {year} {2009})},\ \Eprint {https://arxiv.org/abs/0906.1197} {arXiv:0906.1197 [astro-ph.CO]} \BibitemShut {NoStop}%
\bibitem [{\citenamefont {Madhavacheril}\ \emph {et~al.}(2014)\citenamefont {Madhavacheril}, \citenamefont {Sehgal},\ and\ \citenamefont {Slatyer}}]{Madhavacheril:2013cna}%
  \BibitemOpen
  \bibfield  {author} {\bibinfo {author} {\bibfnamefont {M.~S.}\ \bibnamefont {Madhavacheril}}, \bibinfo {author} {\bibfnamefont {N.}~\bibnamefont {Sehgal}},\ and\ \bibinfo {author} {\bibfnamefont {T.~R.}\ \bibnamefont {Slatyer}},\ }\bibfield  {title} {\bibinfo {title} {{Current Dark Matter Annihilation Constraints from CMB and Low-Redshift Data}},\ }\href {https://doi.org/10.1103/PhysRevD.89.103508} {\bibfield  {journal} {\bibinfo  {journal} {Phys. Rev. D}\ }\textbf {\bibinfo {volume} {89}},\ \bibinfo {pages} {103508} (\bibinfo {year} {2014})},\ \Eprint {https://arxiv.org/abs/1310.3815} {arXiv:1310.3815 [astro-ph.CO]} \BibitemShut {NoStop}%
\bibitem [{\citenamefont {Gonzalez-Morales}\ \emph {et~al.}(2017)\citenamefont {Gonzalez-Morales}, \citenamefont {Profumo},\ and\ \citenamefont {Reynoso-C{\'o}rdova}}]{Gonzalez-Morales:2017jkx}%
  \BibitemOpen
  \bibfield  {author} {\bibinfo {author} {\bibfnamefont {A.~X.}\ \bibnamefont {Gonzalez-Morales}}, \bibinfo {author} {\bibfnamefont {S.}~\bibnamefont {Profumo}},\ and\ \bibinfo {author} {\bibfnamefont {J.}~\bibnamefont {Reynoso-C{\'o}rdova}},\ }\bibfield  {title} {\bibinfo {title} {{Prospects for indirect MeV Dark Matter detection with Gamma Rays in light of Cosmic Microwave Background Constraints}},\ }\href {https://doi.org/10.1103/PhysRevD.96.063520} {\bibfield  {journal} {\bibinfo  {journal} {Phys. Rev. D}\ }\textbf {\bibinfo {volume} {96}},\ \bibinfo {pages} {063520} (\bibinfo {year} {2017})},\ \Eprint {https://arxiv.org/abs/1705.00777} {arXiv:1705.00777 [astro-ph.CO]} \BibitemShut {NoStop}%
\bibitem [{\citenamefont {Sarkar}(1996)}]{Sarkar:1995dd}%
  \BibitemOpen
  \bibfield  {author} {\bibinfo {author} {\bibfnamefont {S.}~\bibnamefont {Sarkar}},\ }\bibfield  {title} {\bibinfo {title} {{Big bang nucleosynthesis and physics beyond the standard model}},\ }\href {https://doi.org/10.1088/0034-4885/59/12/001} {\bibfield  {journal} {\bibinfo  {journal} {Rept. Prog. Phys.}\ }\textbf {\bibinfo {volume} {59}},\ \bibinfo {pages} {1493} (\bibinfo {year} {1996})},\ \Eprint {https://arxiv.org/abs/hep-ph/9602260} {arXiv:hep-ph/9602260} \BibitemShut {NoStop}%
\bibitem [{\citenamefont {Schramm}\ and\ \citenamefont {Turner}(1998)}]{Schramm:1997vs}%
  \BibitemOpen
  \bibfield  {author} {\bibinfo {author} {\bibfnamefont {D.~N.}\ \bibnamefont {Schramm}}\ and\ \bibinfo {author} {\bibfnamefont {M.~S.}\ \bibnamefont {Turner}},\ }\bibfield  {title} {\bibinfo {title} {{Big Bang Nucleosynthesis Enters the Precision Era}},\ }\href {https://doi.org/10.1103/RevModPhys.70.303} {\bibfield  {journal} {\bibinfo  {journal} {Rev. Mod. Phys.}\ }\textbf {\bibinfo {volume} {70}},\ \bibinfo {pages} {303} (\bibinfo {year} {1998})},\ \Eprint {https://arxiv.org/abs/astro-ph/9706069} {arXiv:astro-ph/9706069} \BibitemShut {NoStop}%
\bibitem [{\citenamefont {Jedamzik}(2006)}]{Jedamzik:2006xz}%
  \BibitemOpen
  \bibfield  {author} {\bibinfo {author} {\bibfnamefont {K.}~\bibnamefont {Jedamzik}},\ }\bibfield  {title} {\bibinfo {title} {{Big bang nucleosynthesis constraints on hadronically and electromagnetically decaying relic neutral particles}},\ }\href {https://doi.org/10.1103/PhysRevD.74.103509} {\bibfield  {journal} {\bibinfo  {journal} {Phys. Rev. D}\ }\textbf {\bibinfo {volume} {74}},\ \bibinfo {pages} {103509} (\bibinfo {year} {2006})},\ \Eprint {https://arxiv.org/abs/hep-ph/0604251} {arXiv:hep-ph/0604251} \BibitemShut {NoStop}%
\bibitem [{\citenamefont {Kawasaki}\ \emph {et~al.}(2008)\citenamefont {Kawasaki}, \citenamefont {Kohri}, \citenamefont {Moroi},\ and\ \citenamefont {Yotsuyanagi}}]{Kawasaki:2008qe}%
  \BibitemOpen
  \bibfield  {author} {\bibinfo {author} {\bibfnamefont {M.}~\bibnamefont {Kawasaki}}, \bibinfo {author} {\bibfnamefont {K.}~\bibnamefont {Kohri}}, \bibinfo {author} {\bibfnamefont {T.}~\bibnamefont {Moroi}},\ and\ \bibinfo {author} {\bibfnamefont {A.}~\bibnamefont {Yotsuyanagi}},\ }\bibfield  {title} {\bibinfo {title} {{Big-Bang Nucleosynthesis and Gravitino}},\ }\href {https://doi.org/10.1103/PhysRevD.78.065011} {\bibfield  {journal} {\bibinfo  {journal} {Phys. Rev. D}\ }\textbf {\bibinfo {volume} {78}},\ \bibinfo {pages} {065011} (\bibinfo {year} {2008})},\ \Eprint {https://arxiv.org/abs/0804.3745} {arXiv:0804.3745 [hep-ph]} \BibitemShut {NoStop}%
\bibitem [{\citenamefont {Iocco}\ \emph {et~al.}(2009)\citenamefont {Iocco}, \citenamefont {Mangano}, \citenamefont {Miele}, \citenamefont {Pisanti},\ and\ \citenamefont {Serpico}}]{Iocco:2008va}%
  \BibitemOpen
  \bibfield  {author} {\bibinfo {author} {\bibfnamefont {F.}~\bibnamefont {Iocco}}, \bibinfo {author} {\bibfnamefont {G.}~\bibnamefont {Mangano}}, \bibinfo {author} {\bibfnamefont {G.}~\bibnamefont {Miele}}, \bibinfo {author} {\bibfnamefont {O.}~\bibnamefont {Pisanti}},\ and\ \bibinfo {author} {\bibfnamefont {P.~D.}\ \bibnamefont {Serpico}},\ }\bibfield  {title} {\bibinfo {title} {{Primordial Nucleosynthesis: from precision cosmology to fundamental physics}},\ }\href {https://doi.org/10.1016/j.physrep.2009.02.002} {\bibfield  {journal} {\bibinfo  {journal} {Phys. Rept.}\ }\textbf {\bibinfo {volume} {472}},\ \bibinfo {pages} {1} (\bibinfo {year} {2009})},\ \Eprint {https://arxiv.org/abs/0809.0631} {arXiv:0809.0631 [astro-ph]} \BibitemShut {NoStop}%
\bibitem [{\citenamefont {Pospelov}\ and\ \citenamefont {Pradler}(2010{\natexlab{a}})}]{Pospelov:2010hj}%
  \BibitemOpen
  \bibfield  {author} {\bibinfo {author} {\bibfnamefont {M.}~\bibnamefont {Pospelov}}\ and\ \bibinfo {author} {\bibfnamefont {J.}~\bibnamefont {Pradler}},\ }\bibfield  {title} {\bibinfo {title} {{Big Bang Nucleosynthesis as a Probe of New Physics}},\ }\href {https://doi.org/10.1146/annurev.nucl.012809.104521} {\bibfield  {journal} {\bibinfo  {journal} {Ann. Rev. Nucl. Part. Sci.}\ }\textbf {\bibinfo {volume} {60}},\ \bibinfo {pages} {539} (\bibinfo {year} {2010}{\natexlab{a}})},\ \Eprint {https://arxiv.org/abs/1011.1054} {arXiv:1011.1054 [hep-ph]} \BibitemShut {NoStop}%
\bibitem [{\citenamefont {Fradette}\ and\ \citenamefont {Pospelov}(2017)}]{Fradette:2017sdd}%
  \BibitemOpen
  \bibfield  {author} {\bibinfo {author} {\bibfnamefont {A.}~\bibnamefont {Fradette}}\ and\ \bibinfo {author} {\bibfnamefont {M.}~\bibnamefont {Pospelov}},\ }\bibfield  {title} {\bibinfo {title} {{BBN for the LHC: constraints on lifetimes of the Higgs portal scalars}},\ }\href {https://doi.org/10.1103/PhysRevD.96.075033} {\bibfield  {journal} {\bibinfo  {journal} {Phys. Rev. D}\ }\textbf {\bibinfo {volume} {96}},\ \bibinfo {pages} {075033} (\bibinfo {year} {2017})},\ \Eprint {https://arxiv.org/abs/1706.01920} {arXiv:1706.01920 [hep-ph]} \BibitemShut {NoStop}%
\bibitem [{\citenamefont {Kawasaki}\ \emph {et~al.}(2018)\citenamefont {Kawasaki}, \citenamefont {Kohri}, \citenamefont {Moroi},\ and\ \citenamefont {Takaesu}}]{Kawasaki:2017bqm}%
  \BibitemOpen
  \bibfield  {author} {\bibinfo {author} {\bibfnamefont {M.}~\bibnamefont {Kawasaki}}, \bibinfo {author} {\bibfnamefont {K.}~\bibnamefont {Kohri}}, \bibinfo {author} {\bibfnamefont {T.}~\bibnamefont {Moroi}},\ and\ \bibinfo {author} {\bibfnamefont {Y.}~\bibnamefont {Takaesu}},\ }\bibfield  {title} {\bibinfo {title} {{Revisiting Big-Bang Nucleosynthesis Constraints on Long-Lived Decaying Particles}},\ }\href {https://doi.org/10.1103/PhysRevD.97.023502} {\bibfield  {journal} {\bibinfo  {journal} {Phys. Rev. D}\ }\textbf {\bibinfo {volume} {97}},\ \bibinfo {pages} {023502} (\bibinfo {year} {2018})},\ \Eprint {https://arxiv.org/abs/1709.01211} {arXiv:1709.01211 [hep-ph]} \BibitemShut {NoStop}%
\bibitem [{\citenamefont {Ghosh}(2024)}]{Ghosh:2022frt}%
  \BibitemOpen
  \bibfield  {author} {\bibinfo {author} {\bibfnamefont {D.}~\bibnamefont {Ghosh}},\ }\bibfield  {title} {\bibinfo {title} {{Revisiting big bang nucleosynthesis with a new particle species: effect of co-annihilation with nucleons}},\ }\href {https://doi.org/10.1140/epjc/s10052-024-12963-8} {\bibfield  {journal} {\bibinfo  {journal} {Eur. Phys. J. C}\ }\textbf {\bibinfo {volume} {84}},\ \bibinfo {pages} {623} (\bibinfo {year} {2024})},\ \Eprint {https://arxiv.org/abs/2207.10499} {arXiv:2207.10499 [hep-ph]} \BibitemShut {NoStop}%
\bibitem [{\citenamefont {Dev}\ \emph {et~al.}(2025)\citenamefont {Dev}, \citenamefont {Wu},\ and\ \citenamefont {Xu}}]{Dev:2025pru}%
  \BibitemOpen
  \bibfield  {author} {\bibinfo {author} {\bibfnamefont {P.~S.~B.}\ \bibnamefont {Dev}}, \bibinfo {author} {\bibfnamefont {Q.-f.}\ \bibnamefont {Wu}},\ and\ \bibinfo {author} {\bibfnamefont {X.-J.}\ \bibnamefont {Xu}},\ }\bibfield  {title} {\bibinfo {title} {{No Hiding in the Dark: Cosmological Bounds on Heavy Neutral Leptons with Dark Decay Channels}},\ }\href@noop {} {\  (\bibinfo {year} {2025})},\ \Eprint {https://arxiv.org/abs/2507.12270} {arXiv:2507.12270 [hep-ph]} \BibitemShut {NoStop}%
\bibitem [{\citenamefont {Jedamzik}(2004)}]{Jedamzik:2004er}%
  \BibitemOpen
  \bibfield  {author} {\bibinfo {author} {\bibfnamefont {K.}~\bibnamefont {Jedamzik}},\ }\bibfield  {title} {\bibinfo {title} {{Did something decay, evaporate, or annihilate during Big Bang nucleosynthesis?}},\ }\href {https://doi.org/10.1103/PhysRevD.70.063524} {\bibfield  {journal} {\bibinfo  {journal} {Phys. Rev. D}\ }\textbf {\bibinfo {volume} {70}},\ \bibinfo {pages} {063524} (\bibinfo {year} {2004})},\ \Eprint {https://arxiv.org/abs/astro-ph/0402344} {arXiv:astro-ph/0402344} \BibitemShut {NoStop}%
\bibitem [{\citenamefont {Jedamzik}\ and\ \citenamefont {Pospelov}(2009)}]{Jedamzik:2009uy}%
  \BibitemOpen
  \bibfield  {author} {\bibinfo {author} {\bibfnamefont {K.}~\bibnamefont {Jedamzik}}\ and\ \bibinfo {author} {\bibfnamefont {M.}~\bibnamefont {Pospelov}},\ }\bibfield  {title} {\bibinfo {title} {{Big Bang Nucleosynthesis and Particle Dark Matter}},\ }\href {https://doi.org/10.1088/1367-2630/11/10/105028} {\bibfield  {journal} {\bibinfo  {journal} {New J. Phys.}\ }\textbf {\bibinfo {volume} {11}},\ \bibinfo {pages} {105028} (\bibinfo {year} {2009})},\ \Eprint {https://arxiv.org/abs/0906.2087} {arXiv:0906.2087 [hep-ph]} \BibitemShut {NoStop}%
\bibitem [{\citenamefont {Henning}\ and\ \citenamefont {Murayama}(2012)}]{Henning:2012rm}%
  \BibitemOpen
  \bibfield  {author} {\bibinfo {author} {\bibfnamefont {B.}~\bibnamefont {Henning}}\ and\ \bibinfo {author} {\bibfnamefont {H.}~\bibnamefont {Murayama}},\ }\bibfield  {title} {\bibinfo {title} {{Constraints on Light Dark Matter from Big Bang Nucleosynthesis}},\ }\href@noop {} {\  (\bibinfo {year} {2012})},\ \Eprint {https://arxiv.org/abs/1205.6479} {arXiv:1205.6479 [hep-ph]} \BibitemShut {NoStop}%
\bibitem [{\citenamefont {Depta}\ \emph {et~al.}(2019)\citenamefont {Depta}, \citenamefont {Hufnagel}, \citenamefont {Schmidt-Hoberg},\ and\ \citenamefont {Wild}}]{Depta:2019lbe}%
  \BibitemOpen
  \bibfield  {author} {\bibinfo {author} {\bibfnamefont {P.~F.}\ \bibnamefont {Depta}}, \bibinfo {author} {\bibfnamefont {M.}~\bibnamefont {Hufnagel}}, \bibinfo {author} {\bibfnamefont {K.}~\bibnamefont {Schmidt-Hoberg}},\ and\ \bibinfo {author} {\bibfnamefont {S.}~\bibnamefont {Wild}},\ }\bibfield  {title} {\bibinfo {title} {{BBN constraints on the annihilation of MeV-scale dark matter}},\ }\href {https://doi.org/10.1088/1475-7516/2019/04/029} {\bibfield  {journal} {\bibinfo  {journal} {JCAP}\ }\textbf {\bibinfo {volume} {04}},\ \bibinfo {pages} {029}},\ \Eprint {https://arxiv.org/abs/1901.06944} {arXiv:1901.06944 [hep-ph]} \BibitemShut {NoStop}%
\bibitem [{\citenamefont {Sabti}\ \emph {et~al.}(2020)\citenamefont {Sabti}, \citenamefont {Alvey}, \citenamefont {Escudero}, \citenamefont {Fairbairn},\ and\ \citenamefont {Blas}}]{Sabti:2019mhn}%
  \BibitemOpen
  \bibfield  {author} {\bibinfo {author} {\bibfnamefont {N.}~\bibnamefont {Sabti}}, \bibinfo {author} {\bibfnamefont {J.}~\bibnamefont {Alvey}}, \bibinfo {author} {\bibfnamefont {M.}~\bibnamefont {Escudero}}, \bibinfo {author} {\bibfnamefont {M.}~\bibnamefont {Fairbairn}},\ and\ \bibinfo {author} {\bibfnamefont {D.}~\bibnamefont {Blas}},\ }\bibfield  {title} {\bibinfo {title} {{Refined Bounds on MeV-scale Thermal Dark Sectors from BBN and the CMB}},\ }\href {https://doi.org/10.1088/1475-7516/2020/01/004} {\bibfield  {journal} {\bibinfo  {journal} {JCAP}\ }\textbf {\bibinfo {volume} {01}},\ \bibinfo {pages} {004}},\ \Eprint {https://arxiv.org/abs/1910.01649} {arXiv:1910.01649 [hep-ph]} \BibitemShut {NoStop}%
\bibitem [{\citenamefont {Chu}\ \emph {et~al.}(2022)\citenamefont {Chu}, \citenamefont {Kuo},\ and\ \citenamefont {Pradler}}]{Chu:2022xuh}%
  \BibitemOpen
  \bibfield  {author} {\bibinfo {author} {\bibfnamefont {X.}~\bibnamefont {Chu}}, \bibinfo {author} {\bibfnamefont {J.-L.}\ \bibnamefont {Kuo}},\ and\ \bibinfo {author} {\bibfnamefont {J.}~\bibnamefont {Pradler}},\ }\bibfield  {title} {\bibinfo {title} {{Toward a full description of MeV dark matter decoupling: A self-consistent determination of relic abundance and Neff}},\ }\href {https://doi.org/10.1103/PhysRevD.106.055022} {\bibfield  {journal} {\bibinfo  {journal} {Phys. Rev. D}\ }\textbf {\bibinfo {volume} {106}},\ \bibinfo {pages} {055022} (\bibinfo {year} {2022})},\ \Eprint {https://arxiv.org/abs/2205.05714} {arXiv:2205.05714 [hep-ph]} \BibitemShut {NoStop}%
\bibitem [{\citenamefont {Chu}\ and\ \citenamefont {Pradler}(2024)}]{Chu:2023jyb}%
  \BibitemOpen
  \bibfield  {author} {\bibinfo {author} {\bibfnamefont {X.}~\bibnamefont {Chu}}\ and\ \bibinfo {author} {\bibfnamefont {J.}~\bibnamefont {Pradler}},\ }\bibfield  {title} {\bibinfo {title} {{Minimal mass of thermal dark matter and the viability of millicharged particles affecting 21-cm cosmology}},\ }\href {https://doi.org/10.1103/PhysRevD.109.103510} {\bibfield  {journal} {\bibinfo  {journal} {Phys. Rev. D}\ }\textbf {\bibinfo {volume} {109}},\ \bibinfo {pages} {103510} (\bibinfo {year} {2024})},\ \Eprint {https://arxiv.org/abs/2310.06611} {arXiv:2310.06611 [hep-ph]} \BibitemShut {NoStop}%
\bibitem [{\citenamefont {Pospelov}\ and\ \citenamefont {Pradler}(2010{\natexlab{b}})}]{Pospelov:2010cw}%
  \BibitemOpen
  \bibfield  {author} {\bibinfo {author} {\bibfnamefont {M.}~\bibnamefont {Pospelov}}\ and\ \bibinfo {author} {\bibfnamefont {J.}~\bibnamefont {Pradler}},\ }\bibfield  {title} {\bibinfo {title} {{Metastable GeV-scale particles as a solution to the cosmological lithium problem}},\ }\href {https://doi.org/10.1103/PhysRevD.82.103514} {\bibfield  {journal} {\bibinfo  {journal} {Phys.Rev.}\ }\textbf {\bibinfo {volume} {D82}},\ \bibinfo {pages} {103514} (\bibinfo {year} {2010}{\natexlab{b}})},\ \Eprint {https://arxiv.org/abs/1006.4172} {arXiv:1006.4172 [hep-ph]} \BibitemShut {NoStop}%
%%CITATION = ARXIV:1006.4172;%%
\bibitem [{\citenamefont {McKeen}\ and\ \citenamefont {Omar}(2024)}]{McKeen:2024voa}%
  \BibitemOpen
  \bibfield  {author} {\bibinfo {author} {\bibfnamefont {D.}~\bibnamefont {McKeen}}\ and\ \bibinfo {author} {\bibfnamefont {A.}~\bibnamefont {Omar}},\ }\bibfield  {title} {\bibinfo {title} {{Early dark energy during big bang nucleosynthesis}},\ }\href {https://doi.org/10.1103/PhysRevD.110.103514} {\bibfield  {journal} {\bibinfo  {journal} {Phys. Rev. D}\ }\textbf {\bibinfo {volume} {110}},\ \bibinfo {pages} {103514} (\bibinfo {year} {2024})},\ \Eprint {https://arxiv.org/abs/2407.03508} {arXiv:2407.03508 [hep-ph]} \BibitemShut {NoStop}%
\bibitem [{\citenamefont {Pitrou}\ \emph {et~al.}(2020)\citenamefont {Pitrou}, \citenamefont {Coc}, \citenamefont {Uzan},\ and\ \citenamefont {Vangioni}}]{Pitrou:2019nub}%
  \BibitemOpen
  \bibfield  {author} {\bibinfo {author} {\bibfnamefont {C.}~\bibnamefont {Pitrou}}, \bibinfo {author} {\bibfnamefont {A.}~\bibnamefont {Coc}}, \bibinfo {author} {\bibfnamefont {J.-P.}\ \bibnamefont {Uzan}},\ and\ \bibinfo {author} {\bibfnamefont {E.}~\bibnamefont {Vangioni}},\ }\bibfield  {title} {\bibinfo {title} {{Precision Big Bang Nucleosynthesis with the New Code PRIMAT}},\ }\href {https://doi.org/10.7566/JPSCP.31.011034} {\bibfield  {journal} {\bibinfo  {journal} {JPS Conf. Proc.}\ }\textbf {\bibinfo {volume} {31}},\ \bibinfo {pages} {011034} (\bibinfo {year} {2020})},\ \Eprint {https://arxiv.org/abs/1909.12046} {arXiv:1909.12046 [astro-ph.CO]} \BibitemShut {NoStop}%
\bibitem [{\citenamefont {Burns}\ \emph {et~al.}(2024)\citenamefont {Burns}, \citenamefont {Tait},\ and\ \citenamefont {Valli}}]{Burns:2023sgx}%
  \BibitemOpen
  \bibfield  {author} {\bibinfo {author} {\bibfnamefont {A.-K.}\ \bibnamefont {Burns}}, \bibinfo {author} {\bibfnamefont {T.~M.~P.}\ \bibnamefont {Tait}},\ and\ \bibinfo {author} {\bibfnamefont {M.}~\bibnamefont {Valli}},\ }\bibfield  {title} {\bibinfo {title} {{PRyMordial: the first three minutes, within and beyond the standard model}},\ }\href {https://doi.org/10.1140/epjc/s10052-024-12442-0} {\bibfield  {journal} {\bibinfo  {journal} {Eur. Phys. J. C}\ }\textbf {\bibinfo {volume} {84}},\ \bibinfo {pages} {86} (\bibinfo {year} {2024})},\ \Eprint {https://arxiv.org/abs/2307.07061} {arXiv:2307.07061 [hep-ph]} \BibitemShut {NoStop}%
\bibitem [{\citenamefont {Molaro}\ \emph {et~al.}(2024)\citenamefont {Molaro}, \citenamefont {Bonifacio}, \citenamefont {Cupani},\ and\ \citenamefont {Howk}}]{Molaro:2024wxa}%
  \BibitemOpen
  \bibfield  {author} {\bibinfo {author} {\bibfnamefont {P.}~\bibnamefont {Molaro}}, \bibinfo {author} {\bibfnamefont {P.}~\bibnamefont {Bonifacio}}, \bibinfo {author} {\bibfnamefont {G.}~\bibnamefont {Cupani}},\ and\ \bibinfo {author} {\bibfnamefont {J.~C.}\ \bibnamefont {Howk}},\ }\bibfield  {title} {\bibinfo {title} {{Extragalactic $^{85}$Rb/$^{87}$Rb and $^{6}$Li/$^{7}$Li ratios in the Small Magellanic Cloud}},\ }\href {https://doi.org/10.1051/0004-6361/202449529} {\bibfield  {journal} {\bibinfo  {journal} {Astron. Astrophys.}\ }\textbf {\bibinfo {volume} {690}},\ \bibinfo {pages} {A38} (\bibinfo {year} {2024})},\ \Eprint {https://arxiv.org/abs/2407.10818} {arXiv:2407.10818 [astro-ph.CO]} \BibitemShut {NoStop}%
\bibitem [{\citenamefont {Ryan}\ \emph {et~al.}(2000)\citenamefont {Ryan}, \citenamefont {Beers}, \citenamefont {Olive}, \citenamefont {Fields},\ and\ \citenamefont {Norris}}]{Ryan:1999rxn}%
  \BibitemOpen
  \bibfield  {author} {\bibinfo {author} {\bibfnamefont {S.~G.}\ \bibnamefont {Ryan}}, \bibinfo {author} {\bibfnamefont {T.~C.}\ \bibnamefont {Beers}}, \bibinfo {author} {\bibfnamefont {K.~A.}\ \bibnamefont {Olive}}, \bibinfo {author} {\bibfnamefont {B.~D.}\ \bibnamefont {Fields}},\ and\ \bibinfo {author} {\bibfnamefont {J.~E.}\ \bibnamefont {Norris}},\ }\bibfield  {title} {\bibinfo {title} {{Primordial Lithium and Big Bang Nucleosynthesis}},\ }\href {https://doi.org/10.1086/312492} {\bibfield  {journal} {\bibinfo  {journal} {Astrophys. J. Lett.}\ }\textbf {\bibinfo {volume} {530}},\ \bibinfo {pages} {L57} (\bibinfo {year} {2000})},\ \Eprint {https://arxiv.org/abs/astro-ph/9905211} {arXiv:astro-ph/9905211} \BibitemShut {NoStop}%
\bibitem [{\citenamefont {Olive}\ \emph {et~al.}(2000)\citenamefont {Olive}, \citenamefont {Steigman},\ and\ \citenamefont {Walker}}]{Olive:1999ij}%
  \BibitemOpen
  \bibfield  {author} {\bibinfo {author} {\bibfnamefont {K.~A.}\ \bibnamefont {Olive}}, \bibinfo {author} {\bibfnamefont {G.}~\bibnamefont {Steigman}},\ and\ \bibinfo {author} {\bibfnamefont {T.~P.}\ \bibnamefont {Walker}},\ }\bibfield  {title} {\bibinfo {title} {{Primordial nucleosynthesis: Theory and observations}},\ }\href {https://doi.org/10.1016/S0370-1573(00)00031-4} {\bibfield  {journal} {\bibinfo  {journal} {Phys. Rept.}\ }\textbf {\bibinfo {volume} {333}},\ \bibinfo {pages} {389} (\bibinfo {year} {2000})},\ \Eprint {https://arxiv.org/abs/astro-ph/9905320} {arXiv:astro-ph/9905320} \BibitemShut {NoStop}%
\bibitem [{\citenamefont {Cyburt}\ \emph {et~al.}(2016)\citenamefont {Cyburt}, \citenamefont {Fields}, \citenamefont {Olive},\ and\ \citenamefont {Yeh}}]{RevModPhys.88.015004}%
  \BibitemOpen
  \bibfield  {author} {\bibinfo {author} {\bibfnamefont {R.~H.}\ \bibnamefont {Cyburt}}, \bibinfo {author} {\bibfnamefont {B.~D.}\ \bibnamefont {Fields}}, \bibinfo {author} {\bibfnamefont {K.~A.}\ \bibnamefont {Olive}},\ and\ \bibinfo {author} {\bibfnamefont {T.-H.}\ \bibnamefont {Yeh}},\ }\bibfield  {title} {\bibinfo {title} {Big bang nucleosynthesis: Present status},\ }\href {https://doi.org/10.1103/RevModPhys.88.015004} {\bibfield  {journal} {\bibinfo  {journal} {Rev. Mod. Phys.}\ }\textbf {\bibinfo {volume} {88}},\ \bibinfo {pages} {015004} (\bibinfo {year} {2016})}\BibitemShut {NoStop}%
\bibitem [{\citenamefont {Kolb}\ and\ \citenamefont {Turner}(2019)}]{Kolb:1990vq}%
  \BibitemOpen
  \bibfield  {author} {\bibinfo {author} {\bibfnamefont {E.~W.}\ \bibnamefont {Kolb}}\ and\ \bibinfo {author} {\bibfnamefont {M.~S.}\ \bibnamefont {Turner}},\ }\href {https://doi.org/10.1201/9780429492860} {\emph {\bibinfo {title} {{The Early Universe}}}},\ Vol.~\bibinfo {volume} {69}\ (\bibinfo  {publisher} {Taylor and Francis},\ \bibinfo {year} {2019})\BibitemShut {NoStop}%
\bibitem [{\citenamefont {Copi}\ \emph {et~al.}(1995)\citenamefont {Copi}, \citenamefont {Schramm},\ and\ \citenamefont {Turner}}]{Copi:1994ev}%
  \BibitemOpen
  \bibfield  {author} {\bibinfo {author} {\bibfnamefont {C.~J.}\ \bibnamefont {Copi}}, \bibinfo {author} {\bibfnamefont {D.~N.}\ \bibnamefont {Schramm}},\ and\ \bibinfo {author} {\bibfnamefont {M.~S.}\ \bibnamefont {Turner}},\ }\bibfield  {title} {\bibinfo {title} {{Big bang nucleosynthesis and the baryon density of the universe}},\ }\href {https://doi.org/10.1126/science.7809624} {\bibfield  {journal} {\bibinfo  {journal} {Science}\ }\textbf {\bibinfo {volume} {267}},\ \bibinfo {pages} {192} (\bibinfo {year} {1995})},\ \Eprint {https://arxiv.org/abs/astro-ph/9407006} {arXiv:astro-ph/9407006} \BibitemShut {NoStop}%
\bibitem [{\citenamefont {Aghanim}\ \emph {et~al.}(2020)\citenamefont {Aghanim} \emph {et~al.}}]{Planck:2018vyg}%
  \BibitemOpen
  \bibfield  {author} {\bibinfo {author} {\bibfnamefont {N.}~\bibnamefont {Aghanim}} \emph {et~al.} (\bibinfo {collaboration} {Planck}),\ }\bibfield  {title} {\bibinfo {title} {{Planck 2018 results. VI. Cosmological parameters}},\ }\href {https://doi.org/10.1051/0004-6361/201833910} {\bibfield  {journal} {\bibinfo  {journal} {Astron. Astrophys.}\ }\textbf {\bibinfo {volume} {641}},\ \bibinfo {pages} {A6} (\bibinfo {year} {2020})},\ \bibinfo {note} {[Erratum: Astron.Astrophys. 652, C4 (2021)]},\ \Eprint {https://arxiv.org/abs/1807.06209} {arXiv:1807.06209 [astro-ph.CO]} \BibitemShut {NoStop}%
\bibitem [{\citenamefont {Fields}\ \emph {et~al.}(2020)\citenamefont {Fields}, \citenamefont {Olive}, \citenamefont {Yeh},\ and\ \citenamefont {Young}}]{Fields:2019pfx}%
  \BibitemOpen
  \bibfield  {author} {\bibinfo {author} {\bibfnamefont {B.~D.}\ \bibnamefont {Fields}}, \bibinfo {author} {\bibfnamefont {K.~A.}\ \bibnamefont {Olive}}, \bibinfo {author} {\bibfnamefont {T.-H.}\ \bibnamefont {Yeh}},\ and\ \bibinfo {author} {\bibfnamefont {C.}~\bibnamefont {Young}},\ }\bibfield  {title} {\bibinfo {title} {{Big-Bang Nucleosynthesis after Planck}},\ }\href {https://doi.org/10.1088/1475-7516/2020/03/010} {\bibfield  {journal} {\bibinfo  {journal} {JCAP}\ }\textbf {\bibinfo {volume} {03}},\ \bibinfo {pages} {010}},\ \bibinfo {note} {[Erratum: JCAP 11, E02 (2020)]},\ \Eprint {https://arxiv.org/abs/1912.01132} {arXiv:1912.01132 [astro-ph.CO]} \BibitemShut {NoStop}%
\bibitem [{\citenamefont {Navas}\ \emph {et~al.}(2024)\citenamefont {Navas} \emph {et~al.}}]{ParticleDataGroup:2024cfk}%
  \BibitemOpen
  \bibfield  {author} {\bibinfo {author} {\bibfnamefont {S.}~\bibnamefont {Navas}} \emph {et~al.} (\bibinfo {collaboration} {Particle Data Group}),\ }\bibfield  {title} {\bibinfo {title} {{Review of particle physics}},\ }\href {https://doi.org/10.1103/PhysRevD.110.030001} {\bibfield  {journal} {\bibinfo  {journal} {Phys. Rev. D}\ }\textbf {\bibinfo {volume} {110}},\ \bibinfo {pages} {030001} (\bibinfo {year} {2024})}\BibitemShut {NoStop}%
\bibitem [{\citenamefont {Serpico}\ \emph {et~al.}(2004)\citenamefont {Serpico}, \citenamefont {Esposito}, \citenamefont {Iocco}, \citenamefont {Mangano}, \citenamefont {Miele},\ and\ \citenamefont {Pisanti}}]{Serpico:2004gx}%
  \BibitemOpen
  \bibfield  {author} {\bibinfo {author} {\bibfnamefont {P.~D.}\ \bibnamefont {Serpico}}, \bibinfo {author} {\bibfnamefont {S.}~\bibnamefont {Esposito}}, \bibinfo {author} {\bibfnamefont {F.}~\bibnamefont {Iocco}}, \bibinfo {author} {\bibfnamefont {G.}~\bibnamefont {Mangano}}, \bibinfo {author} {\bibfnamefont {G.}~\bibnamefont {Miele}},\ and\ \bibinfo {author} {\bibfnamefont {O.}~\bibnamefont {Pisanti}},\ }\bibfield  {title} {\bibinfo {title} {{Nuclear reaction network for primordial nucleosynthesis: A Detailed analysis of rates, uncertainties and light nuclei yields}},\ }\href {https://doi.org/10.1088/1475-7516/2004/12/010} {\bibfield  {journal} {\bibinfo  {journal} {JCAP}\ }\textbf {\bibinfo {volume} {12}},\ \bibinfo {pages} {010}},\ \Eprint {https://arxiv.org/abs/astro-ph/0408076} {arXiv:astro-ph/0408076} \BibitemShut {NoStop}%
\bibitem [{\citenamefont {Nagai}\ \emph {et~al.}(2006)\citenamefont {Nagai} \emph {et~al.}}]{Nagai:2006jb}%
  \BibitemOpen
  \bibfield  {author} {\bibinfo {author} {\bibfnamefont {Y.}~\bibnamefont {Nagai}} \emph {et~al.},\ }\bibfield  {title} {\bibinfo {title} {{Measurement of the H-2(n, gamma)H-3 reaction cross section between 10-keV and 550-keV}},\ }\href {https://doi.org/10.1103/PhysRevC.74.025804} {\bibfield  {journal} {\bibinfo  {journal} {Phys. Rev. C}\ }\textbf {\bibinfo {volume} {74}},\ \bibinfo {pages} {025804} (\bibinfo {year} {2006})}\BibitemShut {NoStop}%
\bibitem [{\citenamefont {de~Salas}\ and\ \citenamefont {Pastor}(2016)}]{deSalas:2016ztq}%
  \BibitemOpen
  \bibfield  {author} {\bibinfo {author} {\bibfnamefont {P.~F.}\ \bibnamefont {de~Salas}}\ and\ \bibinfo {author} {\bibfnamefont {S.}~\bibnamefont {Pastor}},\ }\bibfield  {title} {\bibinfo {title} {{Relic neutrino decoupling with flavour oscillations revisited}},\ }\href {https://doi.org/10.1088/1475-7516/2016/07/051} {\bibfield  {journal} {\bibinfo  {journal} {JCAP}\ }\textbf {\bibinfo {volume} {07}},\ \bibinfo {pages} {051}},\ \Eprint {https://arxiv.org/abs/1606.06986} {arXiv:1606.06986 [hep-ph]} \BibitemShut {NoStop}%
\bibitem [{\citenamefont {Escudero}(2019)}]{Escudero:2018mvt}%
  \BibitemOpen
  \bibfield  {author} {\bibinfo {author} {\bibfnamefont {M.}~\bibnamefont {Escudero}},\ }\bibfield  {title} {\bibinfo {title} {{Neutrino decoupling beyond the Standard Model: CMB constraints on the Dark Matter mass with a fast and precise $N_{\rm eff}$ evaluation}},\ }\href {https://doi.org/10.1088/1475-7516/2019/02/007} {\bibfield  {journal} {\bibinfo  {journal} {JCAP}\ }\textbf {\bibinfo {volume} {02}},\ \bibinfo {pages} {007}},\ \Eprint {https://arxiv.org/abs/1812.05605} {arXiv:1812.05605 [hep-ph]} \BibitemShut {NoStop}%
\bibitem [{\citenamefont {Reno}\ and\ \citenamefont {Seckel}(1988)}]{Reno:1987qw}%
  \BibitemOpen
  \bibfield  {author} {\bibinfo {author} {\bibfnamefont {M.~H.}\ \bibnamefont {Reno}}\ and\ \bibinfo {author} {\bibfnamefont {D.}~\bibnamefont {Seckel}},\ }\bibfield  {title} {\bibinfo {title} {{Primordial Nucleosynthesis: The Effects of Injecting Hadrons}},\ }\href {https://doi.org/10.1103/PhysRevD.37.3441} {\bibfield  {journal} {\bibinfo  {journal} {Phys. Rev. D}\ }\textbf {\bibinfo {volume} {37}},\ \bibinfo {pages} {3441} (\bibinfo {year} {1988})}\BibitemShut {NoStop}%
\bibitem [{\citenamefont {Zyla}\ \emph {et~al.}(2020)\citenamefont {Zyla} \emph {et~al.}}]{PDG2020}%
  \BibitemOpen
  \bibfield  {author} {\bibinfo {author} {\bibfnamefont {P.~A.}\ \bibnamefont {Zyla}} \emph {et~al.} (\bibinfo {collaboration} {Particle Data Group}),\ }\bibfield  {title} {\bibinfo {title} {{Review of Particle Physics}},\ }\href {https://doi.org/10.1093/ptep/ptaa104} {\bibfield  {journal} {\bibinfo  {journal} {PTEP}\ }\textbf {\bibinfo {volume} {2020}},\ \bibinfo {pages} {083C01} (\bibinfo {year} {2020})}\BibitemShut {NoStop}%
\bibitem [{\citenamefont {Cooke}\ \emph {et~al.}(2014)\citenamefont {Cooke}, \citenamefont {Pettini}, \citenamefont {Jorgenson}, \citenamefont {Murphy},\ and\ \citenamefont {Steidel}}]{Cooke:2013cba}%
  \BibitemOpen
  \bibfield  {author} {\bibinfo {author} {\bibfnamefont {R.}~\bibnamefont {Cooke}}, \bibinfo {author} {\bibfnamefont {M.}~\bibnamefont {Pettini}}, \bibinfo {author} {\bibfnamefont {R.~A.}\ \bibnamefont {Jorgenson}}, \bibinfo {author} {\bibfnamefont {M.~T.}\ \bibnamefont {Murphy}},\ and\ \bibinfo {author} {\bibfnamefont {C.~C.}\ \bibnamefont {Steidel}},\ }\bibfield  {title} {\bibinfo {title} {{Precision measures of the primordial abundance of deuterium}},\ }\href {https://doi.org/10.1088/0004-637X/781/1/31} {\bibfield  {journal} {\bibinfo  {journal} {Astrophys. J.}\ }\textbf {\bibinfo {volume} {781}},\ \bibinfo {pages} {31} (\bibinfo {year} {2014})},\ \Eprint {https://arxiv.org/abs/1308.3240} {arXiv:1308.3240 [astro-ph.CO]} \BibitemShut {NoStop}%
\bibitem [{\citenamefont {Riemer-S\o{}rensen}\ \emph {et~al.}(2015)\citenamefont {Riemer-S\o{}rensen}, \citenamefont {Webb}, \citenamefont {Crighton}, \citenamefont {Dumont}, \citenamefont {Ali}, \citenamefont {Kotu\v{s}}, \citenamefont {Bainbridge}, \citenamefont {Murphy},\ and\ \citenamefont {Carswell}}]{Riemer-Sorensen:2014aoa}%
  \BibitemOpen
  \bibfield  {author} {\bibinfo {author} {\bibfnamefont {S.}~\bibnamefont {Riemer-S\o{}rensen}}, \bibinfo {author} {\bibfnamefont {J.~K.}\ \bibnamefont {Webb}}, \bibinfo {author} {\bibfnamefont {N.}~\bibnamefont {Crighton}}, \bibinfo {author} {\bibfnamefont {V.}~\bibnamefont {Dumont}}, \bibinfo {author} {\bibfnamefont {K.}~\bibnamefont {Ali}}, \bibinfo {author} {\bibfnamefont {S.}~\bibnamefont {Kotu\v{s}}}, \bibinfo {author} {\bibfnamefont {M.}~\bibnamefont {Bainbridge}}, \bibinfo {author} {\bibfnamefont {M.~T.}\ \bibnamefont {Murphy}},\ and\ \bibinfo {author} {\bibfnamefont {R.}~\bibnamefont {Carswell}},\ }\bibfield  {title} {\bibinfo {title} {{A robust deuterium abundance; Re-measurement of the z=3.256 absorption system towards the quasar PKS1937-1009}},\ }\href {https://doi.org/10.1093/mnras/stu2599} {\bibfield  {journal} {\bibinfo  {journal} {Mon. Not. Roy. Astron. Soc.}\ }\textbf {\bibinfo {volume} {447}},\ \bibinfo {pages} {2925} (\bibinfo {year} {2015})},\ \Eprint {https://arxiv.org/abs/1412.4043}
  {arXiv:1412.4043 [astro-ph.CO]} \BibitemShut {NoStop}%
\bibitem [{\citenamefont {Balashev}\ \emph {et~al.}(2016)\citenamefont {Balashev}, \citenamefont {Zavarygin}, \citenamefont {Ivanchik}, \citenamefont {Telikova},\ and\ \citenamefont {Varshalovich}}]{Balashev:2015hoe}%
  \BibitemOpen
  \bibfield  {author} {\bibinfo {author} {\bibfnamefont {S.~A.}\ \bibnamefont {Balashev}}, \bibinfo {author} {\bibfnamefont {E.~O.}\ \bibnamefont {Zavarygin}}, \bibinfo {author} {\bibfnamefont {A.~V.}\ \bibnamefont {Ivanchik}}, \bibinfo {author} {\bibfnamefont {K.~N.}\ \bibnamefont {Telikova}},\ and\ \bibinfo {author} {\bibfnamefont {D.~A.}\ \bibnamefont {Varshalovich}},\ }\bibfield  {title} {\bibinfo {title} {{The primordial deuterium abundance: subDLA system at $z_{\rm abs}=2.437$ towards the QSO J 1444+2919}},\ }\href {https://doi.org/10.1093/mnras/stw356} {\bibfield  {journal} {\bibinfo  {journal} {Mon. Not. Roy. Astron. Soc.}\ }\textbf {\bibinfo {volume} {458}},\ \bibinfo {pages} {2188} (\bibinfo {year} {2016})},\ \Eprint {https://arxiv.org/abs/1511.01797} {arXiv:1511.01797 [astro-ph.GA]} \BibitemShut {NoStop}%
\bibitem [{\citenamefont {Riemer-S\o{}rensen}\ \emph {et~al.}(2017)\citenamefont {Riemer-S\o{}rensen}, \citenamefont {Kotu\v{s}}, \citenamefont {Webb}, \citenamefont {Ali}, \citenamefont {Dumont}, \citenamefont {Murphy},\ and\ \citenamefont {Carswell}}]{Riemer-Sorensen:2017pey}%
  \BibitemOpen
  \bibfield  {author} {\bibinfo {author} {\bibfnamefont {S.}~\bibnamefont {Riemer-S\o{}rensen}}, \bibinfo {author} {\bibfnamefont {S.}~\bibnamefont {Kotu\v{s}}}, \bibinfo {author} {\bibfnamefont {J.~K.}\ \bibnamefont {Webb}}, \bibinfo {author} {\bibfnamefont {K.}~\bibnamefont {Ali}}, \bibinfo {author} {\bibfnamefont {V.}~\bibnamefont {Dumont}}, \bibinfo {author} {\bibfnamefont {M.~T.}\ \bibnamefont {Murphy}},\ and\ \bibinfo {author} {\bibfnamefont {R.~F.}\ \bibnamefont {Carswell}},\ }\bibfield  {title} {\bibinfo {title} {{A precise deuterium abundance: remeasurement of the z = 3.572 absorption system towards the quasar PKS1937\textendash{}101}},\ }\href {https://doi.org/10.1093/mnras/stx681} {\bibfield  {journal} {\bibinfo  {journal} {Mon. Not. Roy. Astron. Soc.}\ }\textbf {\bibinfo {volume} {468}},\ \bibinfo {pages} {3239} (\bibinfo {year} {2017})},\ \Eprint {https://arxiv.org/abs/1703.06656} {arXiv:1703.06656 [astro-ph.CO]} \BibitemShut {NoStop}%
\bibitem [{\citenamefont {{Zavarygin}}\ \emph {et~al.}(2018)\citenamefont {{Zavarygin}}, \citenamefont {{Webb}}, \citenamefont {{Dumont}},\ and\ \citenamefont {{Riemer-S{\o}rensen}}}]{2018MNRAS}%
  \BibitemOpen
  \bibfield  {author} {\bibinfo {author} {\bibfnamefont {E.~O.}\ \bibnamefont {{Zavarygin}}}, \bibinfo {author} {\bibfnamefont {J.~K.}\ \bibnamefont {{Webb}}}, \bibinfo {author} {\bibfnamefont {V.}~\bibnamefont {{Dumont}}},\ and\ \bibinfo {author} {\bibfnamefont {S.}~\bibnamefont {{Riemer-S{\o}rensen}}},\ }\bibfield  {title} {\bibinfo {title} {{The primordial deuterium abundance at z$_{abs}$ = 2.504 from a high signal-to-noise spectrum of Q1009+2956}},\ }\href {https://doi.org/10.1093/mnras/sty1003} {\bibfield  {journal} {\bibinfo  {journal} {Mon. Not. Roy. Astron. Soc.}\ }\textbf {\bibinfo {volume} {477}},\ \bibinfo {pages} {5536} (\bibinfo {year} {2018})},\ \Eprint {https://arxiv.org/abs/1706.09512} {arXiv:1706.09512 [astro-ph.GA]} \BibitemShut {NoStop}%
\bibitem [{\citenamefont {Cooke}\ \emph {et~al.}(2018)\citenamefont {Cooke}, \citenamefont {Pettini},\ and\ \citenamefont {Steidel}}]{Cooke:2017cwo}%
  \BibitemOpen
  \bibfield  {author} {\bibinfo {author} {\bibfnamefont {R.~J.}\ \bibnamefont {Cooke}}, \bibinfo {author} {\bibfnamefont {M.}~\bibnamefont {Pettini}},\ and\ \bibinfo {author} {\bibfnamefont {C.~C.}\ \bibnamefont {Steidel}},\ }\bibfield  {title} {\bibinfo {title} {{One Percent Determination of the Primordial Deuterium Abundance}},\ }\href {https://doi.org/10.3847/1538-4357/aaab53} {\bibfield  {journal} {\bibinfo  {journal} {Astrophys. J.}\ }\textbf {\bibinfo {volume} {855}},\ \bibinfo {pages} {102} (\bibinfo {year} {2018})},\ \Eprint {https://arxiv.org/abs/1710.11129} {arXiv:1710.11129 [astro-ph.CO]} \BibitemShut {NoStop}%
\bibitem [{\citenamefont {Aver}\ \emph {et~al.}(2021)\citenamefont {Aver}, \citenamefont {Berg}, \citenamefont {Olive}, \citenamefont {Pogge}, \citenamefont {Salzer},\ and\ \citenamefont {Skillman}}]{Aver:2020fon}%
  \BibitemOpen
  \bibfield  {author} {\bibinfo {author} {\bibfnamefont {E.}~\bibnamefont {Aver}}, \bibinfo {author} {\bibfnamefont {D.~A.}\ \bibnamefont {Berg}}, \bibinfo {author} {\bibfnamefont {K.~A.}\ \bibnamefont {Olive}}, \bibinfo {author} {\bibfnamefont {R.~W.}\ \bibnamefont {Pogge}}, \bibinfo {author} {\bibfnamefont {J.~J.}\ \bibnamefont {Salzer}},\ and\ \bibinfo {author} {\bibfnamefont {E.~D.}\ \bibnamefont {Skillman}},\ }\bibfield  {title} {\bibinfo {title} {{Improving helium abundance determinations with Leo P as a case study}},\ }\href {https://doi.org/10.1088/1475-7516/2021/03/027} {\bibfield  {journal} {\bibinfo  {journal} {JCAP}\ }\textbf {\bibinfo {volume} {03}},\ \bibinfo {pages} {027}},\ \Eprint {https://arxiv.org/abs/2010.04180} {arXiv:2010.04180 [astro-ph.CO]} \BibitemShut {NoStop}%
\bibitem [{\citenamefont {Valerdi}\ \emph {et~al.}(2019)\citenamefont {Valerdi}, \citenamefont {Peimbert}, \citenamefont {Peimbert},\ and\ \citenamefont {Sixtos}}]{Valerdi:2019beb}%
  \BibitemOpen
  \bibfield  {author} {\bibinfo {author} {\bibfnamefont {M.}~\bibnamefont {Valerdi}}, \bibinfo {author} {\bibfnamefont {A.}~\bibnamefont {Peimbert}}, \bibinfo {author} {\bibfnamefont {M.}~\bibnamefont {Peimbert}},\ and\ \bibinfo {author} {\bibfnamefont {A.}~\bibnamefont {Sixtos}},\ }\bibfield  {title} {\bibinfo {title} {{Determination of the Primordial Helium Abundance Based on NGC 346, an H ii Region of the Small Magellanic Cloud}},\ }\href {https://doi.org/10.3847/1538-4357/ab14e4} {\bibfield  {journal} {\bibinfo  {journal} {Astrophys. J.}\ }\textbf {\bibinfo {volume} {876}},\ \bibinfo {pages} {98} (\bibinfo {year} {2019})},\ \Eprint {https://arxiv.org/abs/1904.01594} {arXiv:1904.01594 [astro-ph.GA]} \BibitemShut {NoStop}%
\bibitem [{\citenamefont {Fern\'andez}\ \emph {et~al.}(2019)\citenamefont {Fern\'andez}, \citenamefont {Terlevich}, \citenamefont {D\'\i{}az},\ and\ \citenamefont {Terlevich}}]{Fernandez:2019hds}%
  \BibitemOpen
  \bibfield  {author} {\bibinfo {author} {\bibfnamefont {V.}~\bibnamefont {Fern\'andez}}, \bibinfo {author} {\bibfnamefont {E.}~\bibnamefont {Terlevich}}, \bibinfo {author} {\bibfnamefont {A.~I.}\ \bibnamefont {D\'\i{}az}},\ and\ \bibinfo {author} {\bibfnamefont {R.}~\bibnamefont {Terlevich}},\ }\bibfield  {title} {\bibinfo {title} {{A Bayesian direct method implementation to fit emission line spectra: Application to the primordial He abundance determination}},\ }\href {https://doi.org/10.1093/mnras/stz1433} {\bibfield  {journal} {\bibinfo  {journal} {Mon. Not. Roy. Astron. Soc.}\ }\textbf {\bibinfo {volume} {487}},\ \bibinfo {pages} {3221} (\bibinfo {year} {2019})},\ \Eprint {https://arxiv.org/abs/1905.09215} {arXiv:1905.09215 [astro-ph.GA]} \BibitemShut {NoStop}%
\bibitem [{\citenamefont {Kurichin}\ \emph {et~al.}(2021)\citenamefont {Kurichin}, \citenamefont {Kislitsyn}, \citenamefont {Klimenko}, \citenamefont {Balashev},\ and\ \citenamefont {Ivanchik}}]{Kurichin:2021ppm}%
  \BibitemOpen
  \bibfield  {author} {\bibinfo {author} {\bibfnamefont {O.~A.}\ \bibnamefont {Kurichin}}, \bibinfo {author} {\bibfnamefont {P.~A.}\ \bibnamefont {Kislitsyn}}, \bibinfo {author} {\bibfnamefont {V.~V.}\ \bibnamefont {Klimenko}}, \bibinfo {author} {\bibfnamefont {S.~A.}\ \bibnamefont {Balashev}},\ and\ \bibinfo {author} {\bibfnamefont {A.~V.}\ \bibnamefont {Ivanchik}},\ }\bibfield  {title} {\bibinfo {title} {{A new determination of the primordial helium abundance using the analyses of H II region spectra from SDSS}},\ }\href {https://doi.org/10.1093/mnras/stab215} {\bibfield  {journal} {\bibinfo  {journal} {Mon. Not. Roy. Astron. Soc.}\ }\textbf {\bibinfo {volume} {502}},\ \bibinfo {pages} {3045} (\bibinfo {year} {2021})},\ \Eprint {https://arxiv.org/abs/2101.09127} {arXiv:2101.09127 [astro-ph.CO]} \BibitemShut {NoStop}%
\bibitem [{\citenamefont {{Hsyu}}\ \emph {et~al.}(2020)\citenamefont {{Hsyu}}, \citenamefont {{Cooke}}, \citenamefont {{Prochaska}},\ and\ \citenamefont {{Bolte}}}]{2020ApJ}%
  \BibitemOpen
  \bibfield  {author} {\bibinfo {author} {\bibfnamefont {T.}~\bibnamefont {{Hsyu}}}, \bibinfo {author} {\bibfnamefont {R.~J.}\ \bibnamefont {{Cooke}}}, \bibinfo {author} {\bibfnamefont {J.~X.}\ \bibnamefont {{Prochaska}}},\ and\ \bibinfo {author} {\bibfnamefont {M.}~\bibnamefont {{Bolte}}},\ }\bibfield  {title} {\bibinfo {title} {{The PHLEK Survey: A New Determination of the Primordial Helium Abundance}},\ }\href {https://doi.org/10.3847/1538-4357/ab91af} {\bibfield  {journal} {\bibinfo  {journal} {\apj}\ }\textbf {\bibinfo {volume} {896}},\ \bibinfo {eid} {77} (\bibinfo {year} {2020})},\ \Eprint {https://arxiv.org/abs/2005.12290} {arXiv:2005.12290 [astro-ph.GA]} \BibitemShut {NoStop}%
\bibitem [{\citenamefont {{Valerdi}}\ \emph {et~al.}(2021)\citenamefont {{Valerdi}}, \citenamefont {{Peimbert}},\ and\ \citenamefont {{Peimbert}}}]{2021MNRAS}%
  \BibitemOpen
  \bibfield  {author} {\bibinfo {author} {\bibfnamefont {M.}~\bibnamefont {{Valerdi}}}, \bibinfo {author} {\bibfnamefont {A.}~\bibnamefont {{Peimbert}}},\ and\ \bibinfo {author} {\bibfnamefont {M.}~\bibnamefont {{Peimbert}}},\ }\bibfield  {title} {\bibinfo {title} {{Chemical abundances in seven metal-poor H II regions and a determination of the primordial helium abundance}},\ }\href {https://doi.org/10.1093/mnras/stab1543} {\bibfield  {journal} {\bibinfo  {journal} {Mon. Not. Roy. Astron. Soc.}\ }\textbf {\bibinfo {volume} {505}},\ \bibinfo {pages} {3624} (\bibinfo {year} {2021})},\ \Eprint {https://arxiv.org/abs/2105.12260} {arXiv:2105.12260 [astro-ph.GA]} \BibitemShut {NoStop}%
\bibitem [{\citenamefont {{Aver}}\ \emph {et~al.}(2022)\citenamefont {{Aver}}, \citenamefont {{Berg}}, \citenamefont {{Hirschauer}}, \citenamefont {{Olive}}, \citenamefont {{Pogge}}, \citenamefont {{Rogers}}, \citenamefont {{Salzer}},\ and\ \citenamefont {{Skillman}}}]{2022MNRAS}%
  \BibitemOpen
  \bibfield  {author} {\bibinfo {author} {\bibfnamefont {E.}~\bibnamefont {{Aver}}}, \bibinfo {author} {\bibfnamefont {D.~A.}\ \bibnamefont {{Berg}}}, \bibinfo {author} {\bibfnamefont {A.~S.}\ \bibnamefont {{Hirschauer}}}, \bibinfo {author} {\bibfnamefont {K.~A.}\ \bibnamefont {{Olive}}}, \bibinfo {author} {\bibfnamefont {R.~W.}\ \bibnamefont {{Pogge}}}, \bibinfo {author} {\bibfnamefont {N.~S.~J.}\ \bibnamefont {{Rogers}}}, \bibinfo {author} {\bibfnamefont {J.~J.}\ \bibnamefont {{Salzer}}},\ and\ \bibinfo {author} {\bibfnamefont {E.~D.}\ \bibnamefont {{Skillman}}},\ }\bibfield  {title} {\bibinfo {title} {{A comprehensive chemical abundance analysis of the extremely metal poor Leoncino Dwarf galaxy (AGC 198691)}},\ }\href {https://doi.org/10.1093/mnras/stab3226} {\bibfield  {journal} {\bibinfo  {journal} {Mon. Not. Roy. Astron. Soc.}\ }\textbf {\bibinfo {volume} {510}},\ \bibinfo {pages} {373} (\bibinfo {year} {2022})},\ \Eprint {https://arxiv.org/abs/2109.00178} {arXiv:2109.00178 [astro-ph.GA]} \BibitemShut
  {NoStop}%
\bibitem [{\citenamefont {Boudaud}\ \emph {et~al.}(2019)\citenamefont {Boudaud}, \citenamefont {Lacroix}, \citenamefont {Stref},\ and\ \citenamefont {Lavalle}}]{Boudaud:2018oya}%
  \BibitemOpen
  \bibfield  {author} {\bibinfo {author} {\bibfnamefont {M.}~\bibnamefont {Boudaud}}, \bibinfo {author} {\bibfnamefont {T.}~\bibnamefont {Lacroix}}, \bibinfo {author} {\bibfnamefont {M.}~\bibnamefont {Stref}},\ and\ \bibinfo {author} {\bibfnamefont {J.}~\bibnamefont {Lavalle}},\ }\bibfield  {title} {\bibinfo {title} {{Robust cosmic-ray constraints on $p$-wave annihilating MeV dark matter}},\ }\href {https://doi.org/10.1103/PhysRevD.99.061302} {\bibfield  {journal} {\bibinfo  {journal} {Phys. Rev. D}\ }\textbf {\bibinfo {volume} {99}},\ \bibinfo {pages} {061302} (\bibinfo {year} {2019})},\ \Eprint {https://arxiv.org/abs/1810.01680} {arXiv:1810.01680 [astro-ph.HE]} \BibitemShut {NoStop}%
\bibitem [{\citenamefont {Holdom}(1986)}]{Holdom:1985ag}%
  \BibitemOpen
  \bibfield  {author} {\bibinfo {author} {\bibfnamefont {B.}~\bibnamefont {Holdom}},\ }\bibfield  {title} {\bibinfo {title} {{Two U(1)'s and Epsilon Charge Shifts}},\ }\href {https://doi.org/10.1016/0370-2693(86)91377-8} {\bibfield  {journal} {\bibinfo  {journal} {Phys.Lett.}\ }\textbf {\bibinfo {volume} {B166}},\ \bibinfo {pages} {196} (\bibinfo {year} {1986})}\BibitemShut {NoStop}%
%%CITATION = PHLTA,B166,196;%%
\bibitem [{\citenamefont {deNiverville}\ \emph {et~al.}(2012)\citenamefont {deNiverville}, \citenamefont {McKeen},\ and\ \citenamefont {Ritz}}]{deNiverville:2012ij}%
  \BibitemOpen
  \bibfield  {author} {\bibinfo {author} {\bibfnamefont {P.}~\bibnamefont {deNiverville}}, \bibinfo {author} {\bibfnamefont {D.}~\bibnamefont {McKeen}},\ and\ \bibinfo {author} {\bibfnamefont {A.}~\bibnamefont {Ritz}},\ }\bibfield  {title} {\bibinfo {title} {{Signatures of sub-GeV dark matter beams at neutrino experiments}},\ }\href {https://doi.org/10.1103/PhysRevD.86.035022} {\bibfield  {journal} {\bibinfo  {journal} {Phys.Rev.}\ }\textbf {\bibinfo {volume} {D86}},\ \bibinfo {pages} {035022} (\bibinfo {year} {2012})},\ \Eprint {https://arxiv.org/abs/1205.3499} {arXiv:1205.3499 [hep-ph]} \BibitemShut {NoStop}%
%%CITATION = ARXIV:1205.3499;%%
\bibitem [{\citenamefont {Boehm}\ and\ \citenamefont {Fayet}(2004)}]{Boehm:2003hm}%
  \BibitemOpen
  \bibfield  {author} {\bibinfo {author} {\bibfnamefont {C.}~\bibnamefont {Boehm}}\ and\ \bibinfo {author} {\bibfnamefont {P.}~\bibnamefont {Fayet}},\ }\bibfield  {title} {\bibinfo {title} {{Scalar dark matter candidates}},\ }\href {https://doi.org/10.1016/j.nuclphysb.2004.01.015} {\bibfield  {journal} {\bibinfo  {journal} {Nucl.Phys.}\ }\textbf {\bibinfo {volume} {B683}},\ \bibinfo {pages} {219} (\bibinfo {year} {2004})},\ \Eprint {https://arxiv.org/abs/hep-ph/0305261} {arXiv:hep-ph/0305261 [hep-ph]} \BibitemShut {NoStop}%
%%CITATION = HEP-PH/0305261;%%
\bibitem [{\citenamefont {Batell}\ \emph {et~al.}(2009)\citenamefont {Batell}, \citenamefont {Pospelov},\ and\ \citenamefont {Ritz}}]{Batell:2009yf}%
  \BibitemOpen
  \bibfield  {author} {\bibinfo {author} {\bibfnamefont {B.}~\bibnamefont {Batell}}, \bibinfo {author} {\bibfnamefont {M.}~\bibnamefont {Pospelov}},\ and\ \bibinfo {author} {\bibfnamefont {A.}~\bibnamefont {Ritz}},\ }\bibfield  {title} {\bibinfo {title} {{Probing a Secluded U(1) at B-factories}},\ }\href {https://doi.org/10.1103/PhysRevD.79.115008} {\bibfield  {journal} {\bibinfo  {journal} {Phys. Rev. D}\ }\textbf {\bibinfo {volume} {79}},\ \bibinfo {pages} {115008} (\bibinfo {year} {2009})},\ \Eprint {https://arxiv.org/abs/0903.0363} {arXiv:0903.0363 [hep-ph]} \BibitemShut {NoStop}%
\bibitem [{\citenamefont {Lees}\ \emph {et~al.}(2017)\citenamefont {Lees} \emph {et~al.}}]{BaBar:2017tiz}%
  \BibitemOpen
  \bibfield  {author} {\bibinfo {author} {\bibfnamefont {J.~P.}\ \bibnamefont {Lees}} \emph {et~al.} (\bibinfo {collaboration} {BaBar}),\ }\bibfield  {title} {\bibinfo {title} {{Search for Invisible Decays of a Dark Photon Produced in ${e}^{+}{e}^{-}$ Collisions at BaBar}},\ }\href {https://doi.org/10.1103/PhysRevLett.119.131804} {\bibfield  {journal} {\bibinfo  {journal} {Phys. Rev. Lett.}\ }\textbf {\bibinfo {volume} {119}},\ \bibinfo {pages} {131804} (\bibinfo {year} {2017})},\ \Eprint {https://arxiv.org/abs/1702.03327} {arXiv:1702.03327 [hep-ex]} \BibitemShut {NoStop}%
\bibitem [{\citenamefont {Andreev}\ \emph {et~al.}(2023)\citenamefont {Andreev} \emph {et~al.}}]{NA64:2023wbi}%
  \BibitemOpen
  \bibfield  {author} {\bibinfo {author} {\bibfnamefont {Y.~M.}\ \bibnamefont {Andreev}} \emph {et~al.} (\bibinfo {collaboration} {NA64}),\ }\bibfield  {title} {\bibinfo {title} {{Search for Light Dark Matter with NA64 at CERN}},\ }\href {https://doi.org/10.1103/PhysRevLett.131.161801} {\bibfield  {journal} {\bibinfo  {journal} {Phys. Rev. Lett.}\ }\textbf {\bibinfo {volume} {131}},\ \bibinfo {pages} {161801} (\bibinfo {year} {2023})},\ \Eprint {https://arxiv.org/abs/2307.02404} {arXiv:2307.02404 [hep-ex]} \BibitemShut {NoStop}%
\bibitem [{\citenamefont {Angloher}\ \emph {et~al.}(2016)\citenamefont {Angloher} \emph {et~al.}}]{CRESST:2015txj}%
  \BibitemOpen
  \bibfield  {author} {\bibinfo {author} {\bibfnamefont {G.}~\bibnamefont {Angloher}} \emph {et~al.} (\bibinfo {collaboration} {CRESST}),\ }\bibfield  {title} {\bibinfo {title} {{Results on light dark matter particles with a low-threshold CRESST-II detector}},\ }\href {https://doi.org/10.1140/epjc/s10052-016-3877-3} {\bibfield  {journal} {\bibinfo  {journal} {Eur. Phys. J. C}\ }\textbf {\bibinfo {volume} {76}},\ \bibinfo {pages} {25} (\bibinfo {year} {2016})},\ \Eprint {https://arxiv.org/abs/1509.01515} {arXiv:1509.01515 [astro-ph.CO]} \BibitemShut {NoStop}%
\bibitem [{\citenamefont {Berlin}\ \emph {et~al.}(2019)\citenamefont {Berlin}, \citenamefont {Blinov}, \citenamefont {Krnjaic}, \citenamefont {Schuster},\ and\ \citenamefont {Toro}}]{Berlin:2018bsc}%
  \BibitemOpen
  \bibfield  {author} {\bibinfo {author} {\bibfnamefont {A.}~\bibnamefont {Berlin}}, \bibinfo {author} {\bibfnamefont {N.}~\bibnamefont {Blinov}}, \bibinfo {author} {\bibfnamefont {G.}~\bibnamefont {Krnjaic}}, \bibinfo {author} {\bibfnamefont {P.}~\bibnamefont {Schuster}},\ and\ \bibinfo {author} {\bibfnamefont {N.}~\bibnamefont {Toro}},\ }\bibfield  {title} {\bibinfo {title} {{Dark Matter, Millicharges, Axion and Scalar Particles, Gauge Bosons, and Other New Physics with LDMX}},\ }\href {https://doi.org/10.1103/PhysRevD.99.075001} {\bibfield  {journal} {\bibinfo  {journal} {Phys. Rev. D}\ }\textbf {\bibinfo {volume} {99}},\ \bibinfo {pages} {075001} (\bibinfo {year} {2019})},\ \Eprint {https://arxiv.org/abs/1807.01730} {arXiv:1807.01730 [hep-ph]} \BibitemShut {NoStop}%
\end{thebibliography}%
\end{document}